\UseRawInputEncoding

\documentclass[nonatbib,12pt]{elsarticle}
\usepackage{textcomp}
\usepackage{hyperref}
\usepackage{multirow}
\usepackage{makecell}
\usepackage{graphicx}
\usepackage{array}

\journal{Patterns}

\begin{document}

\begin{frontmatter}

\title{Starting a Synthetic Biological Intelligence Lab from Scratch}


\author[1,2]{Md Sayed Tanveer}
\author[1,2,3]{Dhruvik Patel}
\author[5]{Hunter E. Schweiger}
\author[6,7]{Kwaku Dad Abu-Bonsrah}
\author[6]{Brad Watmuff}
\author[6]{Azin Azadi}
\author[2]{Sergey Pryshchep}
\author[2]{Karthikeyan Narayanan}
\author[1,2]{Christopher Puleo}
\author[4]{Kannathal Natarajan}
\author[5]{Mohammed A. Mostajo-Radji\corref{cor1}}
\ead{mmostajo@ucsc.edu}
\author[6]{Brett J. Kagan\corref{cor1}}
\ead{brett@corticallabs.com}
\author[1,2]{Ge Wang\corref{cor2}}
\ead{wangg6@rpi.edu}
\cortext[cor1]{Co-corresponding author}
\cortext[cor2]{Corresponding author}
\affiliation[1]{organization={Department of Biomedical Engineering, Rensselaer Polytechnic Institute}, city={Troy}, state={NY}, postcode={12180}, country={USA}}
\affiliation[2]{organization={Center for Biotechnology and Interdisciplinary Studies, Rensselaer Polytechnic Institute}, city={Troy}, state={NY}, postcode={12180}, country={USA}}
\affiliation[3]{organization={Albany Medical College}, city={Albany}, state={NY}, postcode={12208}, country={USA}}
\affiliation[4]{organization={The Design Lab at Rensselaer, School of Engineering, Rensselaer Polytechnic Institute}, city={Troy}, state={NY}, postcode={12180}, country={USA}}
\affiliation[5]{organization={Genomics Institute, University of California Santa Cruz}, city={Santa Cruz}, state={CA}, postcode={95060}, country={USA}}
\affiliation[6]{organization={Cortical Labs}, city={Melbourne}, state={VIC}, country={Australia}}
\affiliation[7]{organization={Department of Paediatrics, University of Melbourne}, city={Melbourne}, state={VIC}, country={Australia}}

\begin{abstract}
\noindent With the recent advancements in artificial intelligence, researchers and industries are deploying gigantic models trained on billions of samples. While training these models consumes a huge amount of energy, human brains produce similar outputs (along with other capabilities) with massively lower data and energy requirements. For this reason, more researchers are increasingly considering alternatives. One of these alternatives is known as synthetic biological intelligence, which involves training \textit{in vitro} neurons for goal-directed tasks. This multidisciplinary field requires knowledge of tissue engineering, bio-materials, digital signal processing, computer programming, neuroscience, and even artificial intelligence. The multidisciplinary requirements make starting synthetic biological intelligence research highly non-trivial and time-consuming. Generally, most labs either specialize in the biological aspects or the computational ones. Here, we propose how a lab focusing on computational aspects, including machine learning and device interfacing, can start working on synthetic biological intelligence, including organoid intelligence. We will also discuss computational aspects, which can be helpful for labs that focus on biological research. To facilitate synthetic biological intelligence research, we will describe such a general process step by step, including risks and precautions that could lead to substantial delay or additional cost.
\end{abstract}

\begin{keyword}
synthetic biological intelligence \sep organoid intelligence \sep electrophysiology \sep biocomputing \sep neuroscience \sep artificial intelligence
\end{keyword}

\end{frontmatter}


\section*{The Bigger Picture}\label{big_picture}
With the advent of large language models (LLMs) and multimodal multitask models, silicon-based computers are becoming closer to human-level performance for increasingly more tasks. However, these artificial intelligence (AI) models are highly energy-consuming. Biological neural networks, on the other hand, can learn to perform goal-oriented tasks with minimum time and data samples, consuming only a tiny fraction of the energy required by silicon-based artificial neural networks. Recently, interest is rapidly growing in the training of synthetic biological neural networks for the development of NeuroAI, a combination of neuroscience and AI. Research in this direction will help pave the way for artificial general intelligence (AGI). However, for beginners this new and multidisciplinary field can be rather challenging to enter. To bridge this gap, our tutorial sheds light on developing a laboratory for NeuroAI research from scratch and focuses on two of the most important factors of synthetic bio-intelligence: the first is to grow the neural networks, and the second is to interface it electrophysiologically with recording and stimulation devices for synthetic neuron-computer interaction.

\section{Introduction}\label{intro}

Synthetic biological intelligence (SBI) aims to assess the functional information processing capabilities of \textit{in vitro} biological neural networks (BNN) via electrophysiological activity to elicit emergent intelligent behaviors \cite{kagan_technology_2023}. Organoid intelligence (OI) is a subset of SBI, which focuses on brain organoids, widely believed to recapitulate brain architecture and function better than traditional 2D or 3D multilayer cultures \cite{smirnova_organoid_2023}. The focus on using synthetic biology methods to generate neural cell culture from human induced pluripotent stem cells (hiPSCs) has also attracted interest due to the potential of providing human cell sources to SBI and other applications such as personalized medicine approaches to drug development \cite{wang_modeling_2018, tejavibulya_personalized_2016}. This is especially pertinent as the results of current drug development available in the market cannot effectively address the complexity of diverse diseases and patient heterogeneity.\cite{ahsan_heterogeneity_2020, roden_genetic_2002}. Moreover, SBI and OI are gaining momentum amongst computational researchers as potential alternatives to modern silicon-based computers for key areas not optimized on von Neumann architecture. Although the modern era is experiencing a boom in big data and large AI models, such as large language models (LLMs) like ChatGPT \cite{brown_language_2020}, image generation models such as DALL-E \cite{ramesh_zero-shot_2021}, or video generation models such as SORA \cite{liu_sora_2024}, these models require training and fine-tuning billions of parameters using terabytes of data and training them costs megawatts of energy \cite{samsi_words_2023}. On the other hand, animal brains can do similar tasks with very few examples, consuming even less than one watt \cite{levy_computation_2020}. 

Previously, primary rodent cortical monolayers were used to explore simple directional tasks in 2D environments \cite{bakkum_spatio-temporal_2008} and obstacle avoidance \cite{tessadori_closed-loop_2015}. Evidence that more nuanced computational tasks could be performed, such as blind source separation from a mixture of signals \cite{isomura_cultured_2015} in a manner consistent with the free energy principle (FEP) has also been found \cite{friston_free-energy_2010, isomura_vitro_2018}. Implications from the FEP were also applied as part of a real-time closed-loop electrophysiological setup for  \textit{in vitro} neural networks to assess performance in a simplified Pong gameworld \cite{kagan_vitro_2022, goldwag_dishbrain_2023, isomura_experimental_2023}. The finding that structured information would result in real-time changes to network activity consistent with an otherwise arbitrary goal was further supported by evidence that these changes involved a population-level electrophysiological reorganization to display critical dynamics, believed to be important in information transfer and processing \cite{habibollahi_critical_2023}. Moreover, open-loop systems have also generated interesting results through reservoir computing paradigms for both organoids \cite{cai_brain_2023} and structured monolayers \cite{sumi_biological_2023}.

The success of the above work, along with the enabling work covered in detail elsewhere \cite{kagan_technology_2023}, has spurred international collaborative efforts. The field of OI is one example where, to foster greater cooperation, experts from different fields convened and issued the Baltimore Declaration \cite{hartung_baltimore_2023}. This established an OI community committed to exploring the potential technology in an organized manner while adhering to ethical standards \cite{morales_pantoja_first_2023}. The coordinated efforts also highlight the multidisciplinary nature of SBI, where collaboration among researchers from different areas is mandatory. Concurrently, more emphasis is being put on the open sharing of large-scale neurophysiological data among researchers from the biological, cognitive, behavioral, and systems neuroscience fields, and machine learning and deep learning researchers \cite{koch_next-generation_2022, de_vries_sharing_2023, gillon_open_2024}. In line with this, there are already various groups working on developing cloud-based OI platforms for collaboration, which will improve accessibility and may encourage such data sharing \cite{jordan_open_2024, elliott_internet-connected_2023, voitiuk_feedback-driven_2024, zhang_mind_2024, chong_system_2023}. Cloud-based collaboration seems to be a solution where one lab tackles the biological aspects while others handle the engineering and computational tasks. Although this field is still in its infancy, it is supported by an involved scientific community to enjoy rapid development for its revolutionary goals.

While cloud-based SBI access is promising, previously computational-focused groups may wish to explore the biological components (i.e., wetware) independently or may simply wish to understand the basics of what their cell culture-focused collaborators are doing. In this paper, we focus on providing insights to these groups on the necessary environment and adequate skills involved in the wetware experimental challenges. We aim to not only provide insights for a group with limited to no prior knowledge of cell culture but also describe how the algorithmic and device-interfacing aspects are related.

\section{Processing Units: Neuron Cell Culture}\label{sec:cell_culture}
For bio-intelligence, the main processing units are the interconnected neural cells. Many researchers also call it "wetware" since we grow these processing units in a dish with nutrients, which are primarily in the liquid form \cite{bray2009wetware}. Culturing neural cells is essential to kick-start any bio-intelligence lab and requires both an appropriate workspace for neural culture \cite{mukherjee_establishment_2023} and suitable methods.

\subsection{Workspace for Neural (or Any) Cell Culture}
\label{sec:neuron_culture_aspects}

\subsubsection{Sterile Equipment} Cell culture requires a sterile environment to sustain long-term viability without contamination.  The various possibilities of contamination are discussed in Section \ref{sec:contamination}. It is necessary to minimize the risks of contamination using an isolated cell culture room with sterilizable equipment and proper handling so as not to compromise the sterility of the equipment. To culture neurons and other neural cells effectively, essential equipment includes a biosafety cabinet, incubator with CO\textsubscript{2} supply, centrifuge, water bath, refrigeration and freezer (including cell cryopreservation) units, and microscopes for assessing cell viability. Access to a cell culture core facility is recommended for beginners to utilize shared resources and gain experience. Necessary consumables encompass cell culture media and reagents, sterile pipettes and tips, centrifuge tubes, multiwell plates, Petri dishes, and basic lab supplies like ethanol, bleach, gloves, and lab coats to maintain a sterile environment (see \hyperref[suppinfo:equipment]{S2.1.1} for more details).

\subsubsection{Controlled Environment} 
To culture any cell, we must maintain an environment suitable for cell growth and maintenance \cite{harrison_general_1997}. Cells are fragile and sensitive to changes in the environment. Thus, it is vital to control environmental parameters (temperature, air, and humidity) inside an incubator (see \hyperref[suppinfo:controlled_env]{S2.1.2} for more details). For most labs, using a commercial incubator is sufficient for short-term experimentation. However, extended experiments will benefit from either open- or closed-loop culture media perfusion circuits that can integrate other systems, such as long-term MEA device support,  dynamic media conditioning, and continuous microscope scanning. Integrating a perfusion system with dynamic media conditioning and efficient media exchange capabilities is critical, as it allows for the regular replenishment of nutrients and removal of waste products, which are vital for maintaining cell viability and stability in long-term cultures. Media exchange within the perfusion system helps prevent the accumulation of toxic byproducts and ensures that cells receive a steady supply of fresh nutrients, both of which are crucial for sustaining functional and responsive neural networks. Furthermore, continuous microscope scanning is effective for monitoring structural and morphological changes in real-time, ensuring researchers can capture crucial developments as they occur. These features are likely to be important for SBI research, where sustained intelligent neural activities (which are dependent on structural relationships between the cultured neurons) are required \cite{voitiuk_feedback-driven_2024}. Implementing such advanced systems can thus greatly enhance the quality and depth of SBI studies, providing insights that static or short-term setups may not be able to achieve.

\subsubsection{Primary Culture and Cell Lines}
Before we describe cell culture, we mention how to procure cells. Based on the cell sources, we can broadly classify the neuron culture process into two types: (1) primary neuron culture and (2) cell line-based culture.

\begin{enumerate}
    \item \textit{\underline{Primary neuron culture}:} In primary neuron culture, we directly harvest the neurons from the brain tissue of an organism before plating them \cite{geraghty2014guidelines}. Harvested and dissociated neural cells from C$57$BL/$6$ mice or Sprague-Dawley rats are typically used \cite{engber2011c57, wang_recent_2017} as primary cultures. However, primary neurons can be cultured from a variety of species, including humans, non-human primates, and birds \cite{pollen_establishing_2019, faltin_morphological_1985}. Furthermore, primary tissue strategies also gives the opportunity to culture harvested brain slices, which can be maintained \textit{in vitro} for several days \cite{andrews_mtor_2020}. For a description of important factors involved with efficient primary neuron culture, see \hyperref[suppinfo:primary_neurons]{S2.1.3.1}.
    
    \item \textit{\underline{Differentiation of cell lines}:} Cell lines are derived initially from primary tissue that becomes immortalized or obtains the capability to proliferate indefinitely. This can happen either through spontaneous adjustment to \textit{in vitro} conditions \cite{perillo_spontaneous_2023}  or transformed immortalization through the introduction of genetic modifications \cite{voloshin_practical_2023}. However, continuous cell division of cell lines can still lead to the accumulation of mutations, so the maintenance of high passage number cell lines is not advised without stringent quality control \cite{calles_effects_2006}. Neurons are post-mitotic, which inhibits the possibility of culturing them as a cell line; however, it is possible to establish neuronal cell lines to obtain them in two ways: (1) differentiation of cells obtained from neuronal cancerous tissue to a post-mitotic state or (2) directed differentiation of stem cell lines. Cell lines obtained from Glioblastoma and neuroblastoma (cancer cells) are easy to maintain and can differentiate into electrically mature, though heterogeneous, neurons \cite{wang_neurod4_2023}. Stem cells, such as embryonic and induced pluripotent stem cells (ESCs and iPSCs), offer reproducible cell types, access to genetic modification tools, and disease modeling, with the potential to generate neurons from various brain regions for research. Broadly, pluripotent stem cells can be differentiated into neurons either through chemical reprogramming with small molecules or via genetic manipulation for directed differentiation \cite{zhang_rapid_2013, hu_neural_2010}.  For detailed descriptions of these cell lines, see \hyperref[suppinfo:cell_line_factors]{S2.1.3.2} and for the media compositions for cell lines, see \hyperref[suppinfo:stemcell_medium]{S2.1.4.3}. It must be noted that a wide variety of cell lines are available from vendors such as the American Type Culture Collection (ATCC) for groups interested in procuring cell lines.

\end{enumerate}

\subsubsection{Coating Reagents and Culture Medium}
\label{sec:coating_and_medium}
    Once the cell type is selected, we need to decide where to grow the culture and what to feed the cells. In a natural \textit{in vivo} environment, neurons are embedded within a complex extracellular matrix (ECM), which provides structural and biochemical support for their development, functionality, and survival. This ECM is rich in proteins that play critical roles in cell-ECM adhesion, neurite outgrowth, and synaptic stability \cite{song_crosstalk_2018}. These interactions are crucial for the formation and maintenance of functional neural networks. Mimicking this natural environment \textit{in vitro} is vital for cultivating healthy and physiologically relevant neurons. For \textit{in vitro} 2D culture, both the coated substrate and the liquid culture media provide nutrients and create the mechanical and chemical stimulation required to emulate the natural \textit{in vivo} environment.

\begin{itemize}

    \item \textit{\underline{Coating reagents}:} Since neurons naturally prefer adhering to an ECM surface, we need to modify the surface of the culture wells (i.e. growth substrate) by coating them with adhesion proteins or polymers. These polymers could be biological, such as Laminin, Fibronectin, Collagen, or Matrigel; or synthetic, such as Poly-d-lysine (PDL) hydrobromide, Poly-ethyleneimine (PEI), or Poly-L-ornithine (PLO) \cite{liu_coating_2020, egert_heart_2005, chad_cellular_1991}. The general practice is to first apply a primary coating with a synthetic polymer such as PDL, PEI, or PLO and then apply a secondary coating with a biological polymer such as Laminin or Matrigel. For more details, see \hyperref[suppinfo:coating_reagents]{S2.1.4.1} and \hyperref[suppinfo:coating_factors]{S2.2.2}.

    \item \textit{\underline{Culture medium}:} In addition to the coating, we select a proper culture media for cell survival. While all media types contain sugars (which the cells use to maintain healthy metabolism), buffering chemicals (to maintain a target pH), and osmolites (to maintain an osmotic balance with the intracellular environment), some neuronal cell types require additional protein, growth factor, or metabolic additive constituents to maintain cell health. Although the coating requirements do not vary much for primary and stem cell-based cultures, the culture media requirement can vary a lot, so we focus on both of them separately. For details on media used for primary culture, see \hyperref[suppinfo:primary_medium]{S2.1.4.2} and for media used for culturing stem cells, see \hyperref[suppinfo:stemcell_medium]{S2.1.4.3}.

\end{itemize}

\subsubsection{2D and 3D Culture}
\label{sec:2d_3d_culture}
Bio-intelligence research can be performed with a range of cell culture topology and geometry that can be influenced by the surface type of the culture wells and the adhesive coating used. Here, we highlight some key differences between 2D culture and 3D culture \cite{duval_modeling_2017}.

\begin{itemize}
    \item \textit{\underline{Two dimensional (2D) culture}:} The simplest form of culture would be a monolayer culture, often referred to as 2D, where the cells approximate a single layer of interconnected neurons \cite{verma_animal_2020}. In monolayer cultures, cells attach to the culture well surface via substrate coatings such as PDL and Laminin, as described above.

    \item \textit{\underline{Three dimensional (3D) culture}:} There are many ways to make 3D cultures, but we will focus on two of the main methods. The first is a multilayer culture, where we use a thicker and porous substrate coating for the cells to grow in and form connections \cite{laplaca2010three}. The other option is to grow 3D spherical organoids. To do so, we can culture them in standard wells using suitable protocols to promote organogenesis, \cite{lancaster_organogenesis_2014} or in an ultra-low-attachment well so that the cells do not attach to the well surface but to each other \cite{pollen_establishing_2019}. 3D cultures provide a larger number of cell-cell connections compared to 2D cultures, which is generally preferred for bio-intelligence. However, it can be difficult for cells in 3D culture to receive nutrients. In natural tissue, there is enough vasculature (i.e., capillaries supplying blood to surrounding cells) to support the supply of nutrients. Unfortunately, growing a vascular network in \textit{in vitro} culture can be difficult, as it requires co-culture with endothelial and other supporting cells and extremely precise microenvironmental regulation \cite{matsui_vascularization_2021}. Currently, a feasible option is diffusion, which does not require any vasculature support. However, the nutrients can only penetrate approximately $300 \mu m$; thus, the multilayer cultures should not have a depth greater than  $300 \mu m$ \cite{zhang_vascularized_2021}. Similarly, for the organoids, the radius of the spheroids should not be larger than $300 \mu m$. Otherwise, the cultures will lose viability and become necrotic in the deeper regions (core). 

\end{itemize}

It must be mentioned that organoids mimic the \textit{in vivo} conditions (morphologically and functionally) more closely compared to other cultures. However, for beginners, we recommend starting with 2D monolayer culture, as the protocols required for this type of culture are the simplest.

\subsection{Major Steps Involved in Primary Neural Culture}
\label{sec:culture_protocol}
For primary neural culture, we need to perform four tasks in sequence, namely cleaning the culture wells, coating them with suitable materials, plating the cells, and maintaining the cells with suitable media. There are numerous variations and combinations of the involved steps for different types of cells, and it is not possible to summarize them all here. For example, non-adhering cells do not require any coating step, and proliferating cells require cell transfer from a well or flask at regular intervals. However, for this study, focusing on primary neural culture is sufficient. Now we elaborate on the four tasks: 

\begin{enumerate}
    \item \textit{\underline{Cleaning the Culture Wells}:} In the case of MEA wells, we need to clean them according to the manufacturer's instructions before using them. Since MEAs can generally be used multiple times, cleaning them thoroughly before each use will avoid cross- and microbial contaminants. This step is unnecessary if using single-use MEAs, which are individually sterilized, treated, and then disposed of as is further discussed in \hyperref[suppinfo:cleaning]{S2.2.1}.

    \item \textit{\underline{Coating the Surface with Cell Adhesion Polymers}:} As previously stated in Subsection \ref{sec:coating_and_medium}, it is important to coat the surfaces with cell adhesion proteins or polymers for the cells to attach. If the neural cells manage to attach properly, they will be able to extend dendrites and axons to form a neural network. For coating procedures, see \hyperref[suppinfo:coating_reagents]{S2.1.4.1} and for important factors we need to know about coating, see \hyperref[suppinfo:coating_factors]{S2.2.2}.

    \item \textit{\underline{Plating the cells on the coated surface}:} After we have completed coating our wells, we move on to plating the cells and begin their incubation. However, we also need to consider the necessary prior preparations before we plate the cells, including dissociating cells from their previous storage or culture condition, washing and counting the cells (to ensure cells are plated at the correct density), and finally exchanging or placing the cells in new plating media. For more details, see \hyperref[suppinfo:plating]{S2.2.3}.

    \item \textit{\underline{Maintaining the cells by changing the media}:} After we plate neural cells, we need to change the media regularly for cell maintenance and maturation. For a description of various cultural media used for primary neural culture, see \hyperref[suppinfo:primary_medium]{S2.1.4.2}. For further description of the essential factors related to media change, see \hyperref[suppinfo:feeding]{S2.2.4}.
\end{enumerate}


\subsection{Cell Culture Assessment}\label{sec:cell_assess}
As mentioned in Subsection \ref{sec:culture_protocol}, after plating the cells on DIV $0$, we need to monitor the status of the cells so that we can take necessary measures to maintain cell health and viability. It is highly recommended that the cells be checked regularly under a microscope. Since the cells are microscopic, we need suitable microscopes to observe them and perform viability assessment, which we describe in detail below.

\subsubsection{Microscopy Setup}
A standard brightfield microscope (inverted) can be used with Petri dishes, multi-well plates, and transparent MEA well bottoms such as Cytoview or Lumos MEA plates from Axion Biosystems (Atlanta, GA $30309$, USA; visit \url{https://www.axionbiosystems.com/} for more information) or standard MEAs from Multi Channel Systems (MCS). However, a reflected light microscope with proper illumination is required for MEA wells with opaque bottoms, such as BioCircuit MEA from Axion Biosystems or MaxOne HD-MEA from MaxWell Biosystems. For MaxOne HD-MEA, the manufacturers recommend using a reflected DIC configuration or a fluorescence microscope.

In our studies, for transparent multi-well plates, we used an Olympus CKX41 Inverted Phase Contrast Microscope at $20\times$ and $40\times$ objectives. For the HD-MEA, we used the Olympus IX51 Inverted Microscope and the Olympus IX-71 Inverted Widefield Fluorescence Microscope at $10\times$ and $20\times$ objectives.

\subsubsection{Cell Health and Viability}
\label{sec:cell_health_viability}
There different ways of checking on cell viability.

\begin{itemize}
    \item \textit{\underline{Cell Morphology Assessment}:} One of the simplest ways to tell whether a neuron culture is thriving is to check the cell and network structure. The neurons and their dendrites or axons are visible under a microscope with suitable magnification. If the cells are healthy, they will appear rounded, and the dendrites and axons will look fully extended and connected. Unhealthy cells appear unrounded with abnormal cellular borders, with their dendrites and axons broken (see Subsection \ref{sec:visuals} for visual examples). This could be caused by many reasons, such as mechanical strain and/or chemical imbalance. Mechanical strain can be caused by excessive forceful pipetting or unbalanced osmotic pressure due to miscalculation in the amount of media constituents used. Chemical imbalances might result in media pH fluctuations due to cellular metabolism or water evaporation. Cells might also start dying from cytotoxicity if excess coating agents are not washed away properly, if cell debris from dead or dying cells are not removed, or if we add an excessive amount of supplements, growth factors, or antibiotics.
    
    \item \textit{\underline{Preparing cell viability assay using fluorescence imaging}:} Although we can estimate cell culture health with a standard microscope image, fluorescent staining agents are required to gauge viability quantitatively. To calculate the cell viability (the percentage of cells alive), we perform live-dead cell assays using a cocktail of fluorescent reagents. We can use Calcein AM (Invitrogen\texttrademark \ \#$C3099$) to visualize the live cells \cite{aras_assessment_2008}. Under fluorescent light (Ex: $495$ nm blue light and Em: $515$ nm green fluorescence), live cells will appear green using the FITC filter set. To image dead cells, we can use Propidium iodide (eBioscience\texttrademark, Invitrogen\texttrademark \ \#BMS$500$PI), which binds with the nuclei of the dead cells and makes them appear red under fluorescent light (Ex: $535$ nm green-yellow light and Em: $617$ nm red fluorescence). In addition, cell tracking fluorescent dyes or quantum dots (QDs) (Qtracker\texttrademark, Invitrogen\texttrademark \ \#Q$25021$MP) can be routinely used to monitor cells for both short-term (15min to 1hr) and long-term (1 day to 14 days) \cite{pathak_quantum_2006}. 
    However, performing frequent assessments using fluorescence agents and light is itself cytotoxic, requires us to wash the cells regularly, and can stress the cells. For this reason, we have used fluorescence imaging as a terminal step in this study.

\end{itemize}

\subsection{Cell Culture Contamination}
\label{sec:contamination}
One of the major obstacles in any kind of cell culture is contamination \cite{langdon_cell_2004}. Cell culture-based research can take months or even years. Thus, contamination can cost precious time and a considerable amount of money.

\subsubsection{Contamination Factors and Prevention}
To prevent contamination, we should know the possible sources of the contaminants. We first mention microbial contamination, a type of contamination mainly caused by bacteria, fungi, or yeast, which can enter the culture from the surrounding environment, nonsterile equipment, or nonsterile and expired reagents. Generally, biosafety cabinets are designed to circulate airflow in such a way that nothing airborne can get inside them. However, contamination can still happen in multiple ways, which are described in \hyperref[suppinfo:contamination_reasons]{S2.4}.

Clearly, most contaminations can be prevented by diligence. Even if some microbes slip past our attention, antibiotic and antifungal agents can be added to culture media to reduce the risk of contamination growth and overtaking the culture.

\subsubsection{Cross-contamination}
Cross-contamination, the unintended transfer of cells between cultures, is a frequent issue for beginners using shared biosafety cabinets and incubators in core facilities \cite{nelson-rees_responsibility_2001}. For groups working with specific cell lines, cells and reagents from others pose serious risks. To minimize this, the involved groups should discuss resource allocation, avoiding shared use whenever possible. While extra cleaning can help when sharing a biosafety cabinet, incubators are more problematic, and it is advised to use a separate one. Contamination can also occur through shared pipette tips and tubes, so strict storage, sterilization, disposal of used equipment, and sourcing of new equipment are vital.

\subsection{Logistics}
\label{sec:logistics}
One of the most important things in conducting any biological experiment, not just neuron electrophysiology, is to ensure that we have unexpired reagents ready to use, and that we have the necessary tools or equipment ready. In this case, proper logistics primarily involve procuring and storing reagents based on shelf life and storage conditions, preparing aliquots for greater efficiency (i.e., minimizing freeze-thaw cycles), and maintaining an inventory of necessary apparatus and consumables. For detailed descriptions, see \hyperref[suppinfo:logistics]{S2.5}.

For summarized information on the various reagents used and their preparation time (useful for inventory management and procurement scheduling) and methods, see Table \ref{tab:reagents_used}. For detailed step-by-step instructions, refer to Transnetyx tissue (cell vendors) and MaxWell Biosystems (MEA vendors) MaxOne protocols.

\renewcommand{\arraystretch}{1.5}  

\begin{table}[]
\centering
\resizebox{\textwidth}{!}{%
\begin{tabular}{|c|c|c|c|c|c|}
\hline
\textbf{\begin{tabular}[c]{@{}c@{}}Reagent\\ Name\end{tabular}} & \textbf{Manufacturer} & \textbf{Purpose} & \textbf{\begin{tabular}[c]{@{}c@{}}Time\\ (w.r.to\\ DIV $0$)\end{tabular}} & \textbf{Amount/Concentration} & \textbf{CAT no.} \\ \hline
1\% Tergazyme & \begin{tabular}[c]{@{}c@{}}Electron\\ Microscopy\\ Sciences\end{tabular} & \multirow{2}{*}{Cleaning} & before $-3$ & \begin{tabular}[c]{@{}c@{}}5g in $500$ml or $10$g in\\ 1L of DI water\end{tabular} & $60520$ \\ \cline{1-2} \cline{4-6} 
\multirow{3}{*}{PBS 1X} & \multirow{3}{*}{Corning} &  & before $-2$ & \begin{tabular}[c]{@{}c@{}}any amount required,\\ no dilution/mixing needed\end{tabular} & \multirow{3}{*}{$21040$CV} \\ \cline{3-5}
 &  & Media (BDNF Preparation) & before $-2$ & \begin{tabular}[c]{@{}c@{}}$500$mL for $500$mg or 1L for\\ 1g of BSA to get $0.1$\%\end{tabular} &  \\ \cline{3-5}
 &  & \begin{tabular}[c]{@{}c@{}}Viability assay\\ (terminal step)\end{tabular} & after 14+ & \begin{tabular}[c]{@{}c@{}}1mL for dye prep,\\ $0.6$mL needed\end{tabular} &  \\ \hline
Primocin & InvivoGen & \begin{tabular}[c]{@{}c@{}}Media (Antimicrobial,\\ Maintenance +\\ Maturation)\end{tabular} & before $-1$ & \begin{tabular}[c]{@{}c@{}}$500\mu$L (from 1L vial,\\ add to media)\end{tabular} & ant-pm-1 \\ \hline
\begin{tabular}[c]{@{}c@{}}BDNF\\ Recombinant\\ Protein\end{tabular} & Gibco & \begin{tabular}[c]{@{}c@{}}Media (Growth factor, \\      Maintenance +\\ Maturation)\end{tabular} & before $-1$ & \begin{tabular}[c]{@{}c@{}}$10\mu$g in $20\mu$L DI water for $0.5$mg/mL,\\ further diluted in $20$mL of\\ $0.1$\% BSA, add 2.5mL to 25mL media\end{tabular} & PHC$7074$ \\ \hline
\begin{tabular}[c]{@{}c@{}}Bovine Serum\\ Albumin (BSA)\end{tabular} & \begin{tabular}[c]{@{}c@{}}Fisher\\ BioReagents™\end{tabular} & \begin{tabular}[c]{@{}c@{}}Media (BDNF\\ dilution,\\ Maintenance +\\ Maturation)\end{tabular} & before $-1$ & \begin{tabular}[c]{@{}c@{}}$500$mg or 1g in $500$mL\\ or 1L of PBS for\\ $0.1$\% concentration\end{tabular} & BP671-1 \\ \hline
Poly-D-Lysine & \begin{tabular}[c]{@{}c@{}}Corning\\ Biocat\end{tabular} & \multirow{2}{*}{\begin{tabular}[c]{@{}c@{}}Coating (Primary, \\ done on DIV $0$)\end{tabular}} & before $-1$-$0$ & \begin{tabular}[c]{@{}c@{}}$100$ug/ml ($20$ mg in $200$ ml\\ of 1X Borax Buffer)\end{tabular} & $354210$ \\ \cline{1-2} \cline{4-6} 
\begin{tabular}[c]{@{}c@{}}Borax Buffer\\ $20$X\end{tabular} & \begin{tabular}[c]{@{}c@{}}Thermo\\ Scientific\end{tabular} &  & before $-1$-$0$ & \begin{tabular}[c]{@{}c@{}}5mL diluted in 195 mL\\ DI water to obtain\\ $200$mL 1X Borax Buffer\end{tabular} & 28341 \\ \hline
Laminin & Corning & \multirow{2}{*}{\begin{tabular}[c]{@{}c@{}}Coating (Secondary,\\ done on DIV $0$)\end{tabular}} & before $-1$-$0$ & \begin{tabular}[c]{@{}c@{}}$20$ug/ml (1 mg \\ in $50$ ml of media)\end{tabular} & 354239 \\ \cline{1-2} \cline{4-6} 
\multirow{2}{*}{NbActiv4} & \multirow{2}{*}{\begin{tabular}[c]{@{}c@{}}Transnetyx\\ Tissue\\ (Brainbits LLC)\end{tabular}} &  & before $-1$-$0$ & $50$ ml for 1 mg Laminin & \multirow{2}{*}{NC1477957} \\ \cline{3-5}
 &  & \begin{tabular}[c]{@{}c@{}}Media (Maintenance,\\ DIV $-2$ to 6)\end{tabular} & before $-3$ & \begin{tabular}[c]{@{}c@{}}$0.6$mL/well (pre-culture and initial plating),\\ $0.3$mL/well for media change\end{tabular} &  \\ \hline
Brainphys & \begin{tabular}[c]{@{}c@{}}Stemcell\\ Technologies\end{tabular} & \multirow{4}{*}{\begin{tabular}[c]{@{}c@{}}Media (Maturation,\\ DIV 7+)\end{tabular}} & before 6 & \begin{tabular}[c]{@{}c@{}}474.5mL (from $500$mL,\\ $0.3$mL/well for media change)\end{tabular} & $05790$ \\ \cline{1-2} \cline{4-6} 
B27 Plus & Gibco &  & before 6 & $10$mL (add to Brainphys) & A$3582801$ \\ \cline{1-2} \cline{4-6} 
N2 & Gibco &  & before 6 & 5mL (add to Brainphys) & $17502048$ \\ \cline{1-2} \cline{4-6} 
\begin{tabular}[c]{@{}c@{}}CD Lipid\\ Concentrate\end{tabular} & Gibco &  & before 6 & 5mL (add to Brainphys) & $11905031$ \\ \hline
Calcein AM & Invitrogen & \multirow{2}{*}{\begin{tabular}[c]{@{}c@{}}Viability assay\\ (terminal step)\end{tabular}} & after 14+ & \begin{tabular}[c]{@{}c@{}}1mg/mL DMSO stock, \\ $5\mu$L in 1mL PBS\\ for aprox. 5uM stock\end{tabular} & C$3099$ \\ \cline{1-2} \cline{4-6} 
\begin{tabular}[c]{@{}c@{}}Propidium\\ Iodide (PI)\end{tabular} & \begin{tabular}[c]{@{}c@{}}eBioscience,\\ Invitrogen\end{tabular} &  & after 14+ & \begin{tabular}[c]{@{}c@{}}$20$ug/mL stock, $50\mu$L in 1mL PBS\\ recommended by vendors\end{tabular} & BMS$500$PI \\ \hline
\end{tabular}%
}
\caption{Reagents used in various steps of E18 primary cortical neuron culture. Please refer to the cell and MEA vendor protocols for detailed step-by-step instructions.}
\label{tab:reagents_used}
\end{table}

\subsection{Visual Examples}
\label{sec:visuals}
\begin{figure}
    \centering
    \includegraphics[width=1\linewidth]{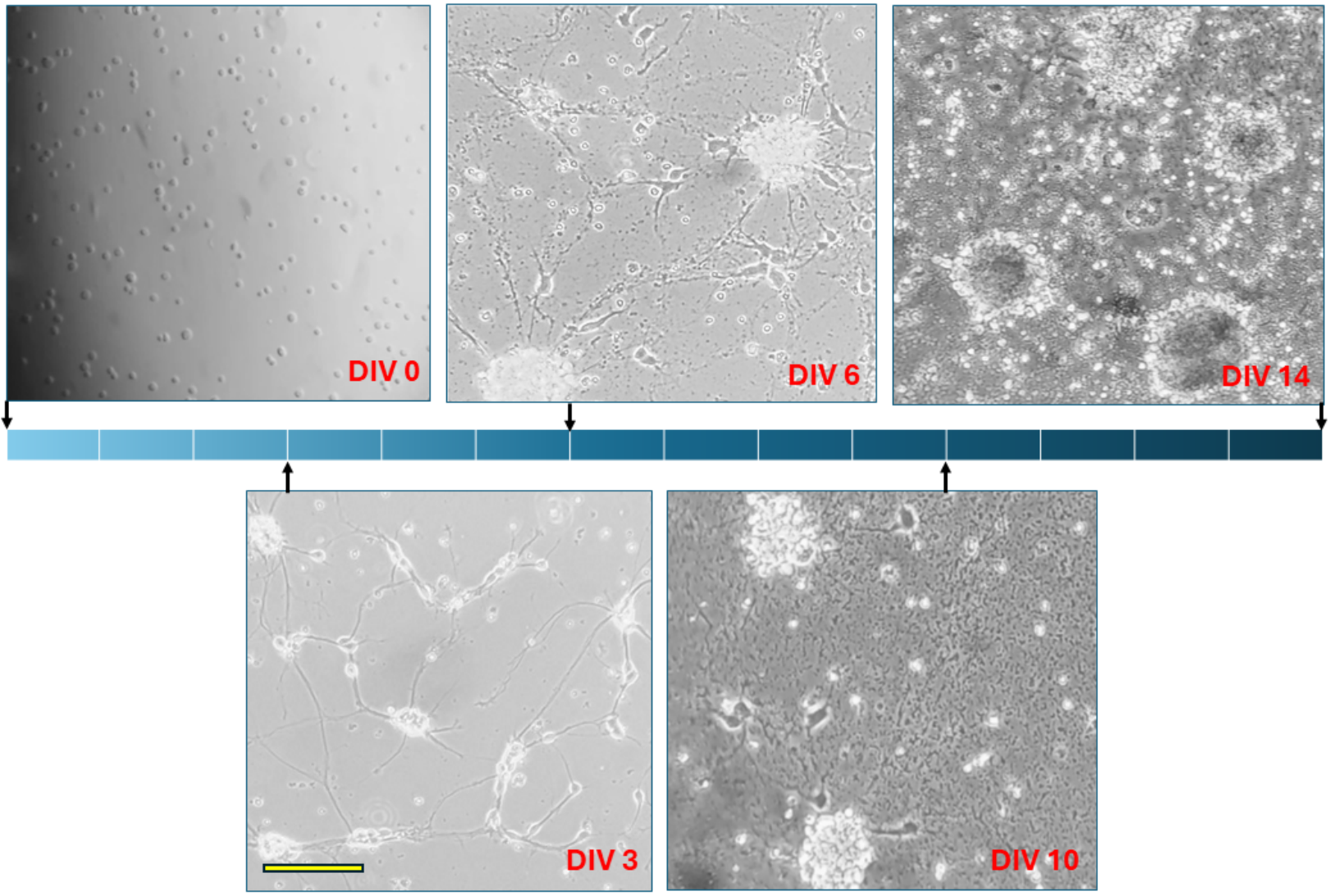}
    \caption{A timeline of how a primary neural cell culture is destroyed due to contamination and other malpractices (scale represents $100\mu$m). While microbes are not apparent at DIV $0$, by DIV 6, they have expanded and are taking over the culture, and by DIV $10$, they have negatively impacted cell viability}
    \label{fig:contamination_example}
\end{figure}

Figure \ref{fig:contamination_example} shows the effects of contamination and improper handling on a cell culture. On DIV $0$, a high number of cells were plated with insufficient coating and without any growth factors or antimicrobial agents. By DIV 3, we can see many cells appearing shriveled and unhealthy in the middle of cell clusters formed by healthier cells connected to one another with axons. However, by DIV 6, the situation has deteriorated significantly: the axons have begun to break down, larger clusters of cells have formed as they struggle to adhere, and a number of microbes are now growing and competing for resources. By DIV $10$, the contaminating microbes have primarily overtaken the culture, with only a few neurons remaining. Finally, by DIV 14, the culture is almost entirely overrun by microbial colonies, with only dead cells remaining.

\begin{figure}
    \centering
    \includegraphics[width=1\linewidth]{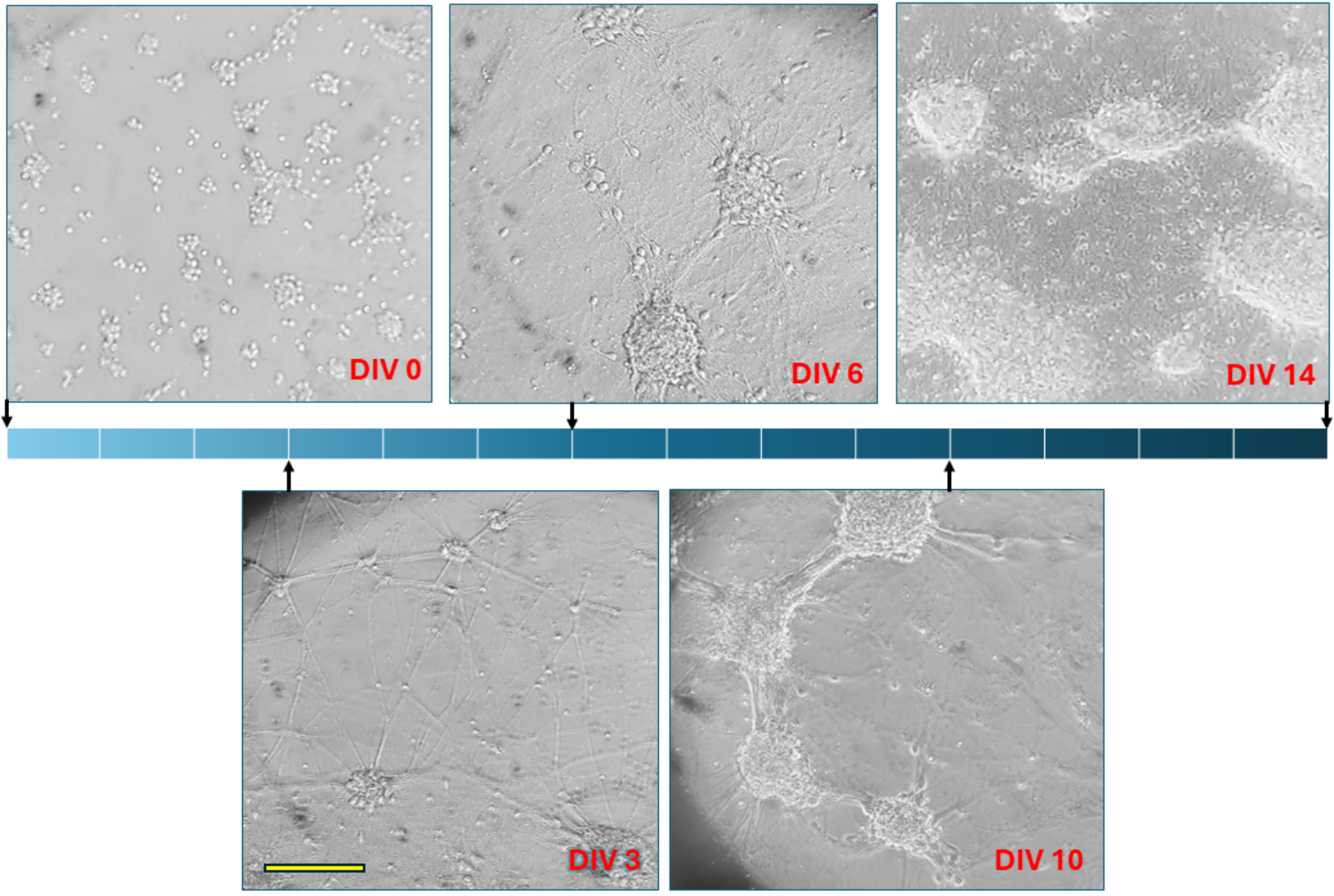}
    \caption{Another timeline of how an adequate primary neural culture could look like with BDNF and Primocin added, and media changed to maturation media on DIV $7$-$10$ (scale represents $100\mu$m). }
    \label{fig:good_enough_culture}
\end{figure}

Figure \ref{fig:good_enough_culture} shows an example of an adequate primary neuron culture. Here, despite a higher cell plating, the dendrites and axons remain extended and healthy even at DIV $10$. On DIV 3, we show a region with smaller clusters to show the healthiness of the axon formations, while on DIV 14, we have shown regions with large clusters that have strong and highly dense axon connections in contrast with the DIV 14 scenario in Figure \ref{fig:contamination_example}.

\begin{figure}
    \centering
    \includegraphics[width=1\linewidth]{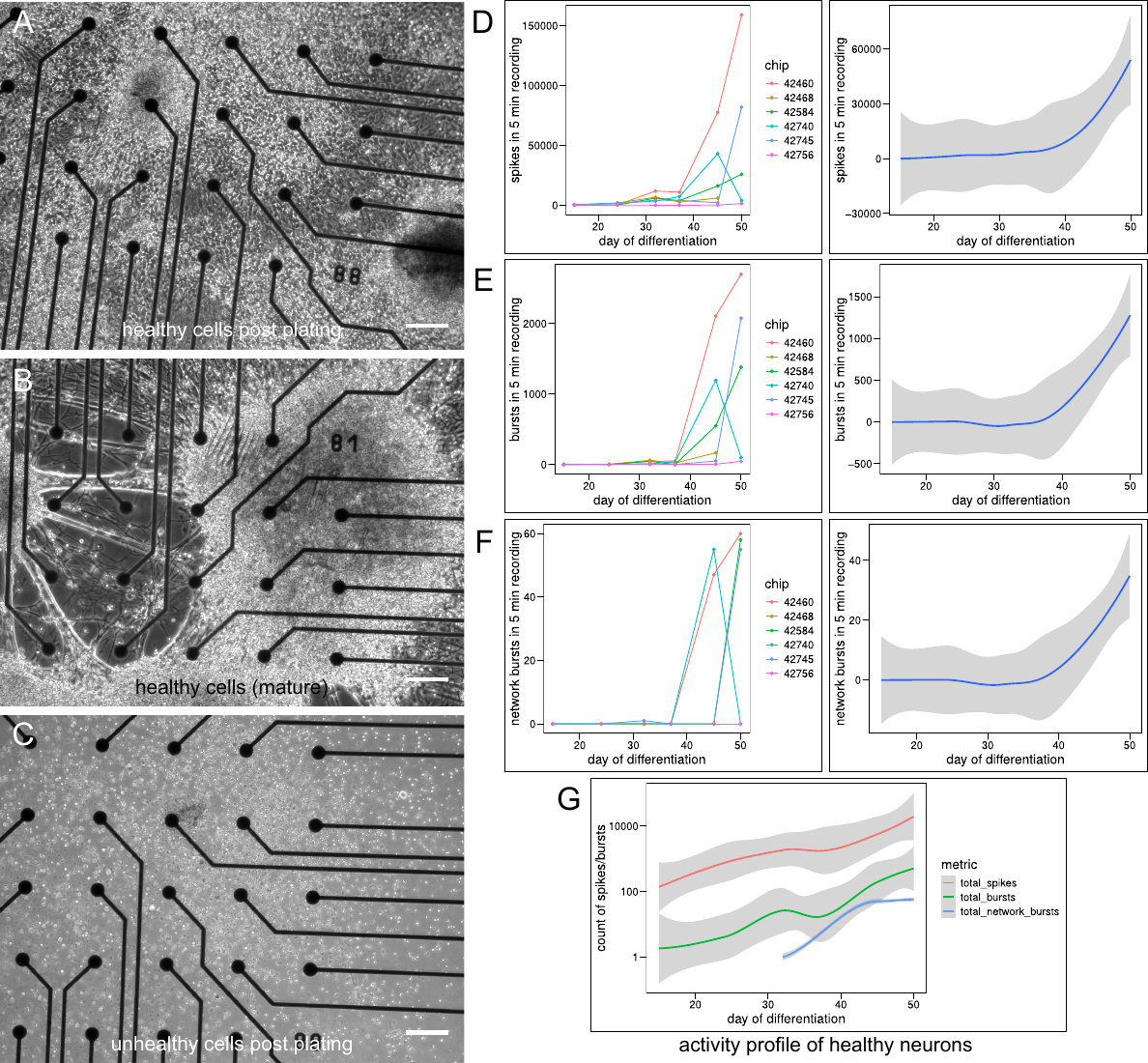}
    \caption{Healthy hiPSC-based neural cells in post-plating (A) and mature (B) states, as well as unhealthy cells (C), grown on a planar-style MEA. (D) to (G)  show various functional metrics contributing to the activity profile. From the top are total spike counts (D), total burst (successive fast spiking) counts (E), and total network burst (successive fast spiking in multiple channels at once) counts (F) over time, with individual MEA chip variability shown in the left panels and regression fit and 95\% CI for trends in the right panels. In (G), the timing of the first appearance of the respective metrics is shown. Scale bar; $100\mu m$.}
    \label{fig:hipsc_example_cortical}
\end{figure}

\begin{figure}
    \centering
    \includegraphics[width=1\linewidth]{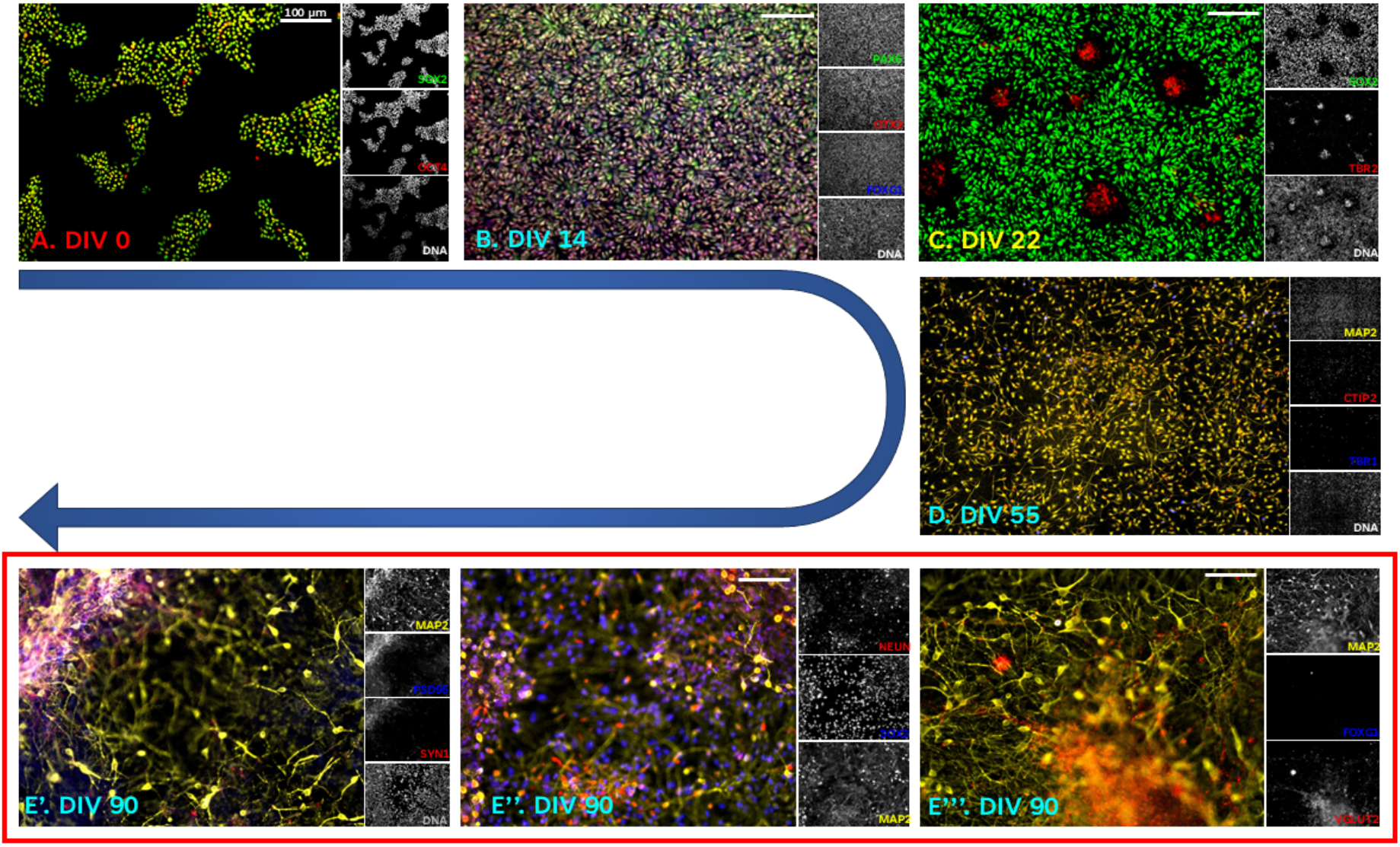}
    \caption{Differentiation of iPSCs to dorsal forebrain progenitors and layer-specific matured cortical projection neurons. Scale bars represent $100\mu$m.\\A: Representative immunofluorescent image showing pluripotency markers (SOX2 and OCT4) of the iPSCs used for the differentiation.\\B: IPSC-derived cortical progenitors self-organize into rosettes and express dorsal (PAX6+) and forebrain (OTX2+) markers at DIV 14.\\C: IPSC-derived cortical progenitors generate Intermediate progenitor cells (IPCs) as marked by TBR2+ at DIV 22.\\D: IPSC-derived cortical progenitors forms both deep-layer cortical projection neurons in long-term culture, DIV 55 as marked by CTIP2 (layer V) and TBR1 (layer VI).\\E: IPSC-derived cortical progenitors mature by expressing synaptic markers, PSD95 and SYN1 (E’), matured markers, NEUN (E’’), and glutamatergic neuronal marker, VGLUT2 (E’’’) at DIV $90$.}
    \label{fig:timeline_stemcell_cortical}
\end{figure}

\begin{figure}
    \centering
    \includegraphics[width=0.75\linewidth]{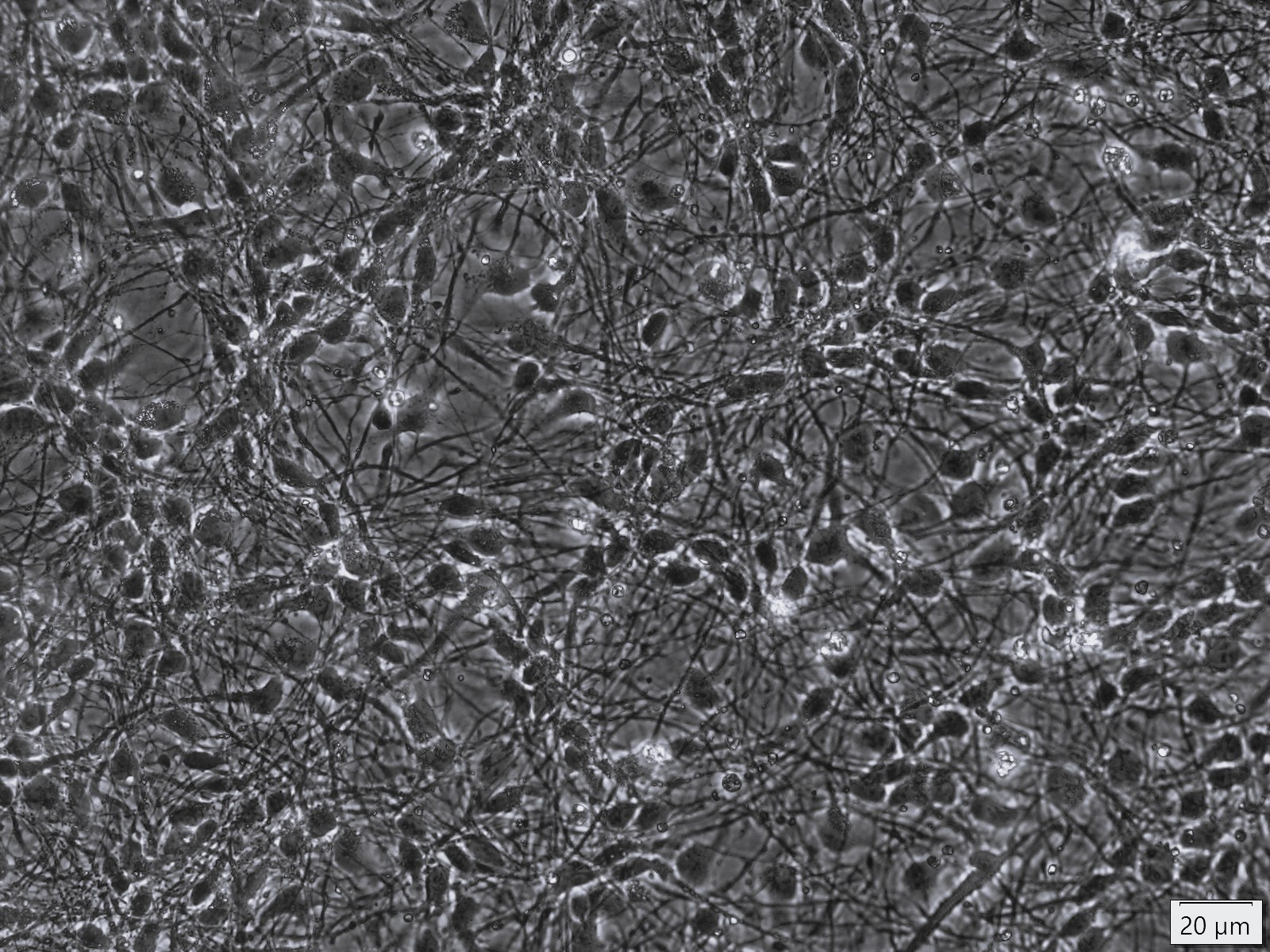}
    \caption{A zoomed-in view of approximately 2D dense neural culture using neural stem cells (scale represents $20\mu$m).}
    \label{fig:ideal_culture}
\end{figure}

Finally, Figure \ref{fig:hipsc_example_cortical} shows a comparative illustration of healthy and unhealthy hiPSC-based neural cells with additional graphs showing spike activity profiles with respect to time, while Figure \ref{fig:timeline_stemcell_cortical} shows the differentiation stages of such cells with respective biomarkers to be expressed in those stages. Here, we can see that for a functional BNN with all feedback mechanisms, we need a dense and mature culture with significant interneuron connections and sufficient biomarker expressions. We show a higher magnification view of such a dense culture in Figure \ref{fig:ideal_culture}, where we can see healthy cell bodies and axon formations. With appropriate training and practice, it is possible to obtain such a lively cell culture.

\section{Electrophysiology and Device Interfacing}\label{electro}
In Section \ref{sec:cell_culture}, we described the different aspects of the neuron culture process. In this section, our primary focus is on recording signals from our grown neuron cells. Since we use primary cortical neurons, as long as we can keep these cells alive for $14$-$28$ days, we can observe complex electrical signals. However, mature population dynamics may take longer to arise. Here, we focus on recording the neuron activities, including spontaneous burst activities and the effects of external stimulations.

\subsection{Different Types of Signal Acquisition Techniques}
To record the electrical activities from neural cells, we require electrodes designed to record and amplify the signals. Neuronal signal acquisition can be classified broadly into two categories: (1) intracellular recording and (2) extracellular recording.

Intracellular recording measures electrical signals from within a neuron, allowing for detailed analysis of membrane potentials. Patch clamp techniques provide precise insights into ion channel activities but are low throughput \cite{nam_vitro_2011}. Calcium imaging, which uses calcium-sensitive dyes or genetically encoded indicators, offers high-throughput population-level recordings but has lower temporal resolution \cite{cameron_prolonged_2017, grodem_updated_2023}. Voltage imaging, using voltage-sensitive dyes or genetically encoded voltage indicators, provides high temporal resolution but a lower signal-to-noise ratio \cite{antic_voltage_2016, adam_voltage_2019}. Extracellular recording involves detecting electric fields outside the cell. Neural probes such as Neuropixel allow recordings from small neuron populations in targeted locations \cite{jun_fully_2017}. MEA wells enable larger network recordings, offering less single-cell resolution but better for overall neuron interactions and long-term studies, making them ideal for SBI development. We have described these techniques in further detail in \hyperref[suppinfo:neuron_recording]{S3.1}.

Considering all options, we recommend beginners start with MEA-based approaches as they are more accessible, typically involve fewer technical hurdles, and require less training to gather initial data.

\subsection{MEA Selection}
Although various types of MEAs have been developed to suit different purposes, most MEA wells can be broadly separated into two classes: passive planar electrode arrays and active high-density complementary metal–oxide–semiconductor (CMOS) arrays. Passive arrays involve metal traces embedded in a substrate and covered with an appropriate insulator, where electrophysiological recording and electrical stimulation are possible. CMOS arrays involve a CMOS chip with multiple electrodes that are sensitive to electrical activity, but that may also be used to provide stimulation to the cells.    
Due to the usability and popularity of MEAs, there are multiple vendors available in the market (we have listed several manufacturers in supplementary section \hyperref[suppinfo:MEA_char]{S3.2}). However, the choice of MEA depends on many factors, and for work on in vitro biological intelligence, the most important factors are yet to be determined. CMOS MEAs provide incredibly high-density recording owing to their small electrode size. They typically have fewer stimulation channels, however, and their smaller electrodes may lead to inconsistencies in impedance and charge delivery, high current density, and, ultimately, a risk of electrolysis. CMOS technology is also not transparent, which makes visualizing cells more challenging. In contrast, passive MEAs typically have far sparser recording electrodes. However, all channels can be used for symmetrical stimulation and reading meaning that more stimulation channels are generally available than with CMOS arrays. The ability to control charge is also typically more nuanced with these MEAs, allowing for potential electrolysis issues to be better controlled. Most planar MEAs are also fabricated on glass and can be visualized. Additionally, the system that controls the MEA is equally important to consider. Currently, no commercially available system designed specifically for real-time closed-loop electrophysiological recording and stimulation is available for purchase. However, many groups are working towards this outcome in different ways \cite{kagan_technology_2023, zhang_mind_2024, jordan_open_2024}.  We have further described the key characteristics to be considered in MEA selection further in \hyperref[suppinfo:MEA_char]{3.2}.

\subsection{Recorded Electric Activities}
\begin{figure}
    \centering
        \includegraphics[width=1\linewidth]{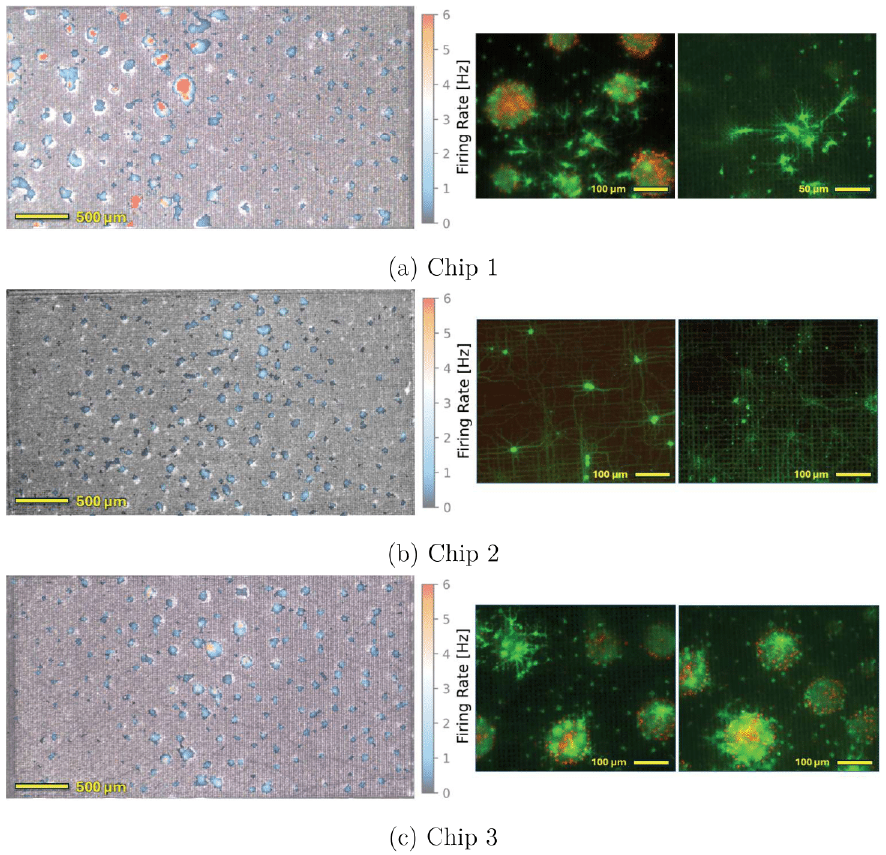}
    \caption{Left: Brightfield images of the three chips, with the electrical activities overlayed. Right: The live and dead cell assays, with green indicating alive cells and red indicating the dead cell nuclei.}
    \label{fig:activity_combined}
\end{figure}

Example recordings and images from both CMOS MaxOne HD-MEAs and on planar MEA are shown in Figure \ref{fig:activity_combined}. HD-MEA was used to record primary neural cultures from mice at DIV 14, where recordings were done using the MaxLab software. Furthermore, planar electrode arrays were used to record human neurons from a hiPSC source, where recordings were done using an in-house MEA system developed by Cortical Labs. In Figure \ref{fig:activity_combined}, the brightfield images are overlaid with the electrical heatmap for HD-MEA chips $1-3$, respectively. While variation exists between chips in cell placement, electrical activities in all three chips suggest that the cell densities in all three chips are in the acceptable range.

\begin{figure}
    \centering
    \includegraphics[width=1\linewidth]{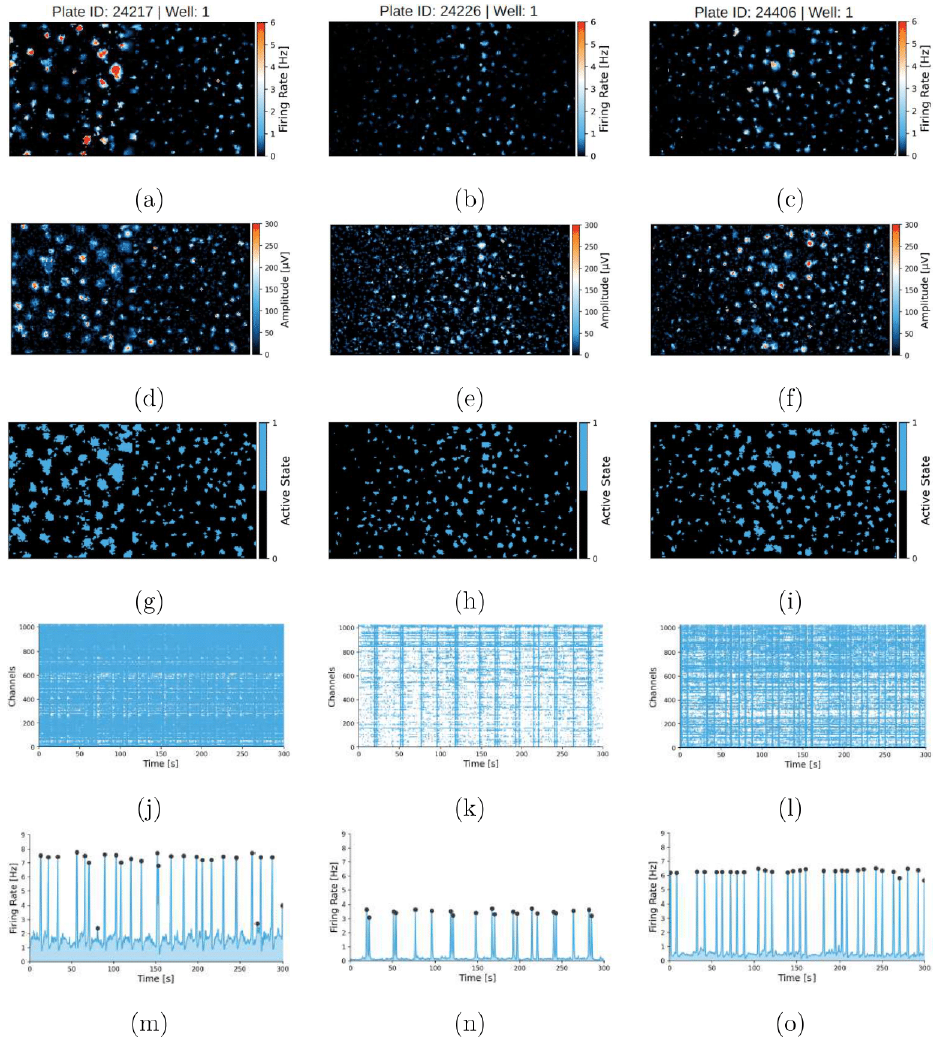}
    \caption{Heatmaps of firing rates, spike amplitudes, active electrodes, channel-wise raster plots, and network activity with respect to time for all three chips. The leftmost, middle, and rightmost columns represent the data from Chip $1-3$, respectively.}
    \label{fig:activity_heatmap_raster}
\end{figure}

We can also see live and dead cell viability assays in the middle and rightmost images in Figure \ref{fig:activity_combined}, where the live cells appear green due to the Calcein AM while the dead cells appear red due to propidium iodide. It should be possible to minimize the eventual death of cells over time as well as their effects on the remaining cells by developing a dedicated setup with a closed-loop environment and automated media control, which we have discussed in Section \ref{sec:disc}. We can also see the heatmaps of the firing rate, spike amplitude, and active electrodes, as well as the channel-wise time domain raster plot and network activity with respect to time in Figure \ref{fig:activity_heatmap_raster}. It is evident that all the chips have a high amount of spontaneous activity, even without any electrical stimulation. However, patterns of cell culture spontaneous activity are variable both within and between cultures and highly dependent on the methods employed. Moreover, patterns of activity will change throughout time, eventually entering quiescence without additional treatment \cite{wagenaar_extremely_2006}. In Figures \ref{fig:activity_heatmap_raster}(d)-(f), the spike amplitudes are within $300 \mu$V, with the majority of them to be within $150 \mu$V. Figures \ref{fig:activity_heatmap_raster}(g)-(i) show the number of spikes recorded only when a burst was detected, while figures \ref{fig:activity_heatmap_raster}(j)-(l) show the time interval recorded between bursts. As seen from the raster plots in Figures \ref{fig:activity_heatmap_raster}(j-l), some synchronized activity can also occur, suggesting some cellular connectivity. In some cases, patterns of activity appear to propagate in a pattern throughout a dish, as seen in Figure \ref{fig:burst_screenshot}. Additional burst activity information is illustrated in supplementary Figures \ref{fig:activity_histograms}.

\subsection{Programming Feedback}
Proper implementation of real-time closed-loop feedback requires the MEA system to be driven by highly efficient low-level hardware and software implementations. Most approaches previously published involve integrations of field-programmable gate arrays (FPGA) that interface between a host computer and the MEA dish and often include dedicated ASIC chips to handle the electrophysiology \cite{muller_sub-millisecond_2012, zhang_mind_2024, newman_closed-loop_2012}. Yet, without a deep technical background in the required programming methods for the different hardware, this approach is not accessible to most scientists. A more accessible approach is to use high-level languages such as Python for high-level decision-making, signal processing, server management, and graphical rendering, as well as low-level data access using C or C++ programming, which Python allows and is built on. While using exclusively low-level languages to minimize any latency in execution would yield good performance, this also requires advanced programming skills with extensive knowledge of computer hardware architecture to achieve and, even then, will result in slower iteration times. The ideal solution is for a system with a flexible low-level setup to be built that allows interactions with a higher-level language that is interpreted to give the performance required with the ease of use and rapid iteration cycles that are desirable. Not surprisingly, the biological and computational sides of SBI and OI have their own challenges. Moreover, even though some hardware and software are available to achieve real-time closed-loop systems, the methods of directing neural cells toward any particular behavior are still under development. For this reason, multidisciplinary researchers should work together to achieve competency and synergy to advance the fields of SBI/OI.

\section{Discussions}\label{sec:disc}
Our initial focus has been on conducting neuron electrophysiology using a minimal setup, given that many new labs lack the resources to build a specialized environment. However, as a lab grows, transitioning to a dedicated setup is essential for ensuring neuron culture longevity, which is crucial for evaluating whether and how SBI (Synthetic Biological Intelligence) and OI (Organoid Intelligence) can become viable alternatives to silicon-based systems. Unlike the human body, \textit{in vitro} neuron cultures are susceptible to contamination and environmental stress, significantly shortening their lifespan. A dedicated setup would include a closed environment with proper temperature and gas control, automated media exchange, and sealed computer connections for long-term signal recording and continuous stimulation. Several platforms, such as UC Santa Cruz’s IoT-enabled microfluidic system \cite{elliott_internet-connected_2023, voitiuk_feedback-driven_2024}, Indiana University’s Brainoware\cite{cai_brain_2023}, Finalspark’s Neuroplatform\cite{jordan_open_2024}, and Cortical Labs CL-1 system \cite{corticallabs1}, aim offer such environments, providing real-time remote access to neural cultures. While differences between each offering exist, it will be crucial for these groups to engage with the scientific community to perform well-validated studies of any equipment made commercially available. 

Stable equipment to support cellular longevity is critical not only for studying neural networks but also for advancing disease modeling. Recent developments favor brain organoids over traditional 2D neuron cultures for modeling complex conditions like Alzheimer’s and Parkinson’s diseases, Autism, and Schizophrenia, as 3D structures mimic the human brain architecture better than their 2D versions \cite{wang_modeling_2018, tejavibulya_personalized_2016, li_organoid_2020}. While many models look into cellular morphology, SBI and OI provide the opportunity to explore functional and even pseudo-cognitive aspects through electrophysiology readouts \cite{haring_microphysiological_2017, habibollahi_critical_2023}. This allows researchers to analyze the neural activity patterns of neural disease models against phenotypically healthy ones, providing deeper insights into how diseases impact brain function. Additionally, these platforms can be used to study the effects of various drugs on brain organoids, offering a comprehensive approach to disease modeling and therapeutic development by addressing both structural and functional aspects of neurological disorders. Thus, building a dedicated setup not only enhances the longevity of neural cultures but also significantly improves the precision and depth of disease modeling. Finally, the approach provides a powerful lens through which to understand the fundamental aspects of how neural systems function. While SBI and OI can provide us with all these benefits and insights, it is also important to consider ethical considerations of growing \textit{in vitro} BNN for embodied experimentation\cite{kagan_scientific_2023, rommelfanger_conceptual_2023, pereira_neural_2023}, along with calls for nomenclature consensus along multiple disciplines to enable better communication \cite{kagan_toward_2024}. However, provided care is taken to fully understand what the ethical implications may arise in this field, the technological improvements can even be leveraged for the purpose of what has been called 'experimental neuroethics'\cite{kagan_embodied_2024}, where ethically relevant metrics are discovered through iterative approaches.

\section{Conclusion}\label{sec:conclusion}
Developing a lab focusing on the multidisciplinary field of SBI/OI can be a daunting undertaking, especially if we lack expertise in either the biological or computational aspects. Thus, we have shared our experience developing a lab that integrates various aspects of SBI and OI. We hope that it will help people from both fields come together and understand each others' importance, as well as work together to make great progress in developing a strong alternative to silicon-based computers. We also hope that more and more groups will join forces in developing SBI and OI for various neuroAI and disease-modeling applications. Overall, we envision a healthy and productive future for all of us and generations to come.

\clearpage
\section*{Acknowledgements}
This work is partially supported by G.W.'s endowment grant, which has been authorized for use by M.S.T and D.P. in this work.
H.E.S. is partially supported by the Graduate Research Fellowship Program of the National Science Foundation.

\section*{Potential Conflict of Interest}
B.J.K., B.W., A.A., and K.D.A. are employees of and may hold shares or another interest in Cortical Labs, a research-focused start-up working in a related space and holding patents related to this manuscript. No specific incentive was provided to any author for contribution to this manuscript.

\section*{Generative AI Statement}
Generative AI, specifically ChatGPT, has been used to check for errors, typos, redundancies, and inconsistencies while writing. However, the article has been written and carefully revised by all the coauthors through multiple iterations.

\section*{Abbreviations}
\textbf{AM}: Acetoxy-Methyl Ester; \textbf{BDNF}: Brain-Derived Neurotrophic Factor; \textbf{BNN}: Biological Neural Network; \textbf{BSA}: Bovine Serum Albumin; \textbf{CMOS}: Complementary Metal-Oxide-Semiconductor; \textbf{EDTA}: Ethylenediaminetetraacetic acid; \textbf{EPSC}: Excitatory Postsynaptic Current; \textbf{ESC}: Embryonic Stem Cells; \textbf{FEP}: Free Energy Principle; \textbf{iPSC}: Induced Pluripotent Stem Cells; \textbf{IPSC}: Inhibitory Postsynaptic Current; \textbf{MEA}: Microelectrode Array; \textbf{OI}: Organoid Intelligence; \textbf{PBS}: Phosphate-Buffered Saline; \textbf{PDL}: Poly-D-Lysine; \textbf{PEI}: Poly-Ethyleneimine; \textbf{PLO}: Poly-L-Ornithine; \textbf{SBI}: Synthetic Biological Intelligence.

\clearpage
\bibliographystyle{elsarticle-num} 
\bibliography{getwriting,references} 

\begin{thebibliography}{100}
\expandafter\ifx\csname url\endcsname\relax
  \def\url#1{\texttt{#1}}\fi
\expandafter\ifx\csname urlprefix\endcsname\relax\def\urlprefix{URL }\fi
\expandafter\ifx\csname href\endcsname\relax
  \def\href#1#2{#2} \def\path#1{#1}\fi

\bibitem{kagan_technology_2023}
B.~J. Kagan, C.~Gyngell, T.~Lysaght, V.~M. Cole, T.~Sawai, J.~Savulescu, \href{https://www.sciencedirect.com/science/article/pii/S0734975023001404}{The technology, opportunities, and challenges of {Synthetic} {Biological} {Intelligence}}, Biotechnology Advances 68 (2023) 108233.
\newblock \href {https://doi.org/10.1016/j.biotechadv.2023.108233} {\path{doi:10.1016/j.biotechadv.2023.108233}}.
\newline\urlprefix\url{https://www.sciencedirect.com/science/article/pii/S0734975023001404}

\bibitem{smirnova_organoid_2023}
L.~Smirnova, B.~S. Caffo, D.~H. Gracias, Q.~Huang, I.~E. Morales~Pantoja, B.~Tang, D.~J. Zack, C.~A. Berlinicke, J.~L. Boyd, T.~D. Harris, E.~C. Johnson, B.~J. Kagan, J.~Kahn, A.~R. Muotri, B.~L. Paulhamus, J.~C. Schwamborn, J.~Plotkin, A.~S. Szalay, J.~T. Vogelstein, P.~F. Worley, T.~Hartung, \href{https://www.frontiersin.org/journals/science/articles/10.3389/fsci.2023.1017235/full}{Organoid intelligence ({OI}): the new frontier in biocomputing and intelligence-in-a-dish}, Frontiers in Science 1, publisher: Frontiers (Feb. 2023).
\newblock \href {https://doi.org/10.3389/fsci.2023.1017235} {\path{doi:10.3389/fsci.2023.1017235}}.
\newline\urlprefix\url{https://www.frontiersin.org/journals/science/articles/10.3389/fsci.2023.1017235/full}

\bibitem{wang_modeling_2018}
H.~Wang, Modeling {Neurological} {Diseases} {With} {Human} {Brain} {Organoids}, Frontiers in Synaptic Neuroscience 10 (2018) 15.
\newblock \href {https://doi.org/10.3389/fnsyn.2018.00015} {\path{doi:10.3389/fnsyn.2018.00015}}.

\bibitem{tejavibulya_personalized_2016}
N.~Tejavibulya, S.~K. Sia, \href{https://www.sciencedirect.com/science/article/pii/S2405471216303660}{Personalized {Disease} {Models} on a {Chip}}, Cell Systems 3~(5) (2016) 416--418.
\newblock \href {https://doi.org/10.1016/j.cels.2016.11.002} {\path{doi:10.1016/j.cels.2016.11.002}}.
\newline\urlprefix\url{https://www.sciencedirect.com/science/article/pii/S2405471216303660}

\bibitem{ahsan_heterogeneity_2020}
T.~Ahsan, N.~J. Urmi, A.~A. Sajib, Heterogeneity in the distribution of 159 drug-response related {SNPs} in world populations and their genetic relatedness, PloS One 15~(1) (2020) e0228000.
\newblock \href {https://doi.org/10.1371/journal.pone.0228000} {\path{doi:10.1371/journal.pone.0228000}}.

\bibitem{roden_genetic_2002}
D.~M. Roden, A.~L. George~Jr, \href{https://www.nature.com/articles/nrd705}{The genetic basis of variability in drug responses}, Nature Reviews Drug Discovery 1~(1) (2002) 37--44, publisher: Nature Publishing Group.
\newblock \href {https://doi.org/10.1038/nrd705} {\path{doi:10.1038/nrd705}}.
\newline\urlprefix\url{https://www.nature.com/articles/nrd705}

\bibitem{brown_language_2020}
T.~Brown, B.~Mann, N.~Ryder, M.~Subbiah, J.~D. Kaplan, P.~Dhariwal, A.~Neelakantan, P.~Shyam, G.~Sastry, A.~Askell, S.~Agarwal, A.~Herbert-Voss, G.~Krueger, T.~Henighan, R.~Child, A.~Ramesh, D.~Ziegler, J.~Wu, C.~Winter, C.~Hesse, M.~Chen, E.~Sigler, M.~Litwin, S.~Gray, B.~Chess, J.~Clark, C.~Berner, S.~McCandlish, A.~Radford, I.~Sutskever, D.~Amodei, \href{https://papers.nips.cc/paper/2020/hash/1457c0d6bfcb4967418bfb8ac142f64a-Abstract.html}{Language {Models} are {Few}-{Shot} {Learners}}, in: Advances in {Neural} {Information} {Processing} {Systems}, Vol.~33, Curran Associates, Inc., 2020, pp. 1877--1901.
\newline\urlprefix\url{https://papers.nips.cc/paper/2020/hash/1457c0d6bfcb4967418bfb8ac142f64a-Abstract.html}

\bibitem{ramesh_zero-shot_2021}
A.~Ramesh, M.~Pavlov, G.~Goh, S.~Gray, C.~Voss, A.~Radford, M.~Chen, I.~Sutskever, \href{https://proceedings.mlr.press/v139/ramesh21a.html}{Zero-{Shot} {Text}-to-{Image} {Generation}}, in: Proceedings of the 38th {International} {Conference} on {Machine} {Learning}, PMLR, 2021, pp. 8821--8831, iSSN: 2640-3498.
\newline\urlprefix\url{https://proceedings.mlr.press/v139/ramesh21a.html}

\bibitem{liu_sora_2024}
Y.~Liu, K.~Zhang, Y.~Li, Z.~Yan, C.~Gao, R.~Chen, Z.~Yuan, Y.~Huang, H.~Sun, J.~Gao, L.~He, L.~Sun, \href{http://arxiv.org/abs/2402.17177}{Sora: {A} {Review} on {Background}, {Technology}, {Limitations}, and {Opportunities} of {Large} {Vision} {Models}}, arXiv:2402.17177 (Apr. 2024).
\newblock \href {https://doi.org/10.48550/arXiv.2402.17177} {\path{doi:10.48550/arXiv.2402.17177}}.
\newline\urlprefix\url{http://arxiv.org/abs/2402.17177}

\bibitem{samsi_words_2023}
S.~Samsi, D.~Zhao, J.~McDonald, B.~Li, A.~Michaleas, M.~Jones, W.~Bergeron, J.~Kepner, D.~Tiwari, V.~Gadepally, \href{https://ieeexplore.ieee.org/abstract/document/10363447}{From {Words} to {Watts}: {Benchmarking} the {Energy} {Costs} of {Large} {Language} {Model} {Inference}}, in: 2023 {IEEE} {High} {Performance} {Extreme} {Computing} {Conference} ({HPEC}), 2023, pp. 1--9, iSSN: 2643-1971.
\newblock \href {https://doi.org/10.1109/HPEC58863.2023.10363447} {\path{doi:10.1109/HPEC58863.2023.10363447}}.
\newline\urlprefix\url{https://ieeexplore.ieee.org/abstract/document/10363447}

\bibitem{levy_computation_2020}
W.~B. Levy, V.~G. Calvert, \href{https://www.biorxiv.org/content/10.1101/2020.04.23.057927v1}{Computation in the human cerebral cortex uses less than 0.2 watts yet this great expense is optimal when considering communication costs}, pages: 2020.04.23.057927 Section: New Results (Apr. 2020).
\newblock \href {https://doi.org/10.1101/2020.04.23.057927} {\path{doi:10.1101/2020.04.23.057927}}.
\newline\urlprefix\url{https://www.biorxiv.org/content/10.1101/2020.04.23.057927v1}

\bibitem{bakkum_spatio-temporal_2008}
D.~J. Bakkum, Z.~C. Chao, S.~M. Potter, Spatio-temporal electrical stimuli shape behavior of an embodied cortical network in a goal-directed learning task, Journal of Neural Engineering 5~(3) (2008) 310--323.
\newblock \href {https://doi.org/10.1088/1741-2560/5/3/004} {\path{doi:10.1088/1741-2560/5/3/004}}.

\bibitem{tessadori_closed-loop_2015}
J.~Tessadori, M.~Chiappalone, Closed-loop neuro-robotic experiments to test computational properties of neuronal networks, Journal of Visualized Experiments: JoVE~(97) (2015) 52341.
\newblock \href {https://doi.org/10.3791/52341} {\path{doi:10.3791/52341}}.

\bibitem{isomura_cultured_2015}
T.~Isomura, K.~Kotani, Y.~Jimbo, \href{https://journals.plos.org/ploscompbiol/article?id=10.1371/journal.pcbi.1004643}{Cultured {Cortical} {Neurons} {Can} {Perform} {Blind} {Source} {Separation} {According} to the {Free}-{Energy} {Principle}}, PLOS Computational Biology 11~(12) (2015) e1004643, publisher: Public Library of Science.
\newblock \href {https://doi.org/10.1371/journal.pcbi.1004643} {\path{doi:10.1371/journal.pcbi.1004643}}.
\newline\urlprefix\url{https://journals.plos.org/ploscompbiol/article?id=10.1371/journal.pcbi.1004643}

\bibitem{friston_free-energy_2010}
K.~Friston, \href{https://www.nature.com/articles/nrn2787}{The free-energy principle: a unified brain theory?}, Nature Reviews Neuroscience 11~(2) (2010) 127--138, publisher: Nature Publishing Group.
\newblock \href {https://doi.org/10.1038/nrn2787} {\path{doi:10.1038/nrn2787}}.
\newline\urlprefix\url{https://www.nature.com/articles/nrn2787}

\bibitem{isomura_vitro_2018}
T.~Isomura, K.~Friston, In vitro neural networks minimise variational free energy, Scientific Reports 8~(1) (2018) 16926.
\newblock \href {https://doi.org/10.1038/s41598-018-35221-w} {\path{doi:10.1038/s41598-018-35221-w}}.

\bibitem{kagan_vitro_2022}
B.~J. Kagan, A.~C. Kitchen, N.~T. Tran, F.~Habibollahi, M.~Khajehnejad, B.~J. Parker, A.~Bhat, B.~Rollo, A.~Razi, K.~J. Friston, In vitro neurons learn and exhibit sentience when embodied in a simulated game-world, Neuron 110~(23) (2022) 3952--3969.e8.
\newblock \href {https://doi.org/10.1016/j.neuron.2022.09.001} {\path{doi:10.1016/j.neuron.2022.09.001}}.

\bibitem{goldwag_dishbrain_2023}
J.~Goldwag, G.~Wang, \href{https://www.nature.com/articles/s42256-023-00666-w}{{DishBrain} plays {Pong} and promises more}, Nature Machine Intelligence 5~(6) (2023) 568--569, publisher: Nature Publishing Group.
\newblock \href {https://doi.org/10.1038/s42256-023-00666-w} {\path{doi:10.1038/s42256-023-00666-w}}.
\newline\urlprefix\url{https://www.nature.com/articles/s42256-023-00666-w}

\bibitem{isomura_experimental_2023}
T.~Isomura, K.~Kotani, Y.~Jimbo, K.~J. Friston, \href{https://www.nature.com/articles/s41467-023-40141-z}{Experimental validation of the free-energy principle with in vitro neural networks}, Nature Communications 14~(1) (2023) 4547, publisher: Nature Publishing Group.
\newblock \href {https://doi.org/10.1038/s41467-023-40141-z} {\path{doi:10.1038/s41467-023-40141-z}}.
\newline\urlprefix\url{https://www.nature.com/articles/s41467-023-40141-z}

\bibitem{habibollahi_critical_2023}
F.~Habibollahi, B.~J. Kagan, A.~N. Burkitt, C.~French, \href{https://www.nature.com/articles/s41467-023-41020-3}{Critical dynamics arise during structured information presentation within embodied in vitro neuronal networks}, Nature Communications 14~(1) (2023) 5287, publisher: Nature Publishing Group.
\newblock \href {https://doi.org/10.1038/s41467-023-41020-3} {\path{doi:10.1038/s41467-023-41020-3}}.
\newline\urlprefix\url{https://www.nature.com/articles/s41467-023-41020-3}

\bibitem{cai_brain_2023}
H.~Cai, Z.~Ao, C.~Tian, Z.~Wu, H.~Liu, J.~Tchieu, M.~Gu, K.~Mackie, F.~Guo, \href{https://www.nature.com/articles/s41928-023-01069-w}{Brain organoid reservoir computing for artificial intelligence}, Nature Electronics 6~(12) (2023) 1032--1039, publisher: Nature Publishing Group.
\newblock \href {https://doi.org/10.1038/s41928-023-01069-w} {\path{doi:10.1038/s41928-023-01069-w}}.
\newline\urlprefix\url{https://www.nature.com/articles/s41928-023-01069-w}

\bibitem{sumi_biological_2023}
T.~Sumi, H.~Yamamoto, Y.~Katori, K.~Ito, S.~Moriya, T.~Konno, S.~Sato, A.~Hirano-Iwata, \href{https://www.pnas.org/doi/10.1073/pnas.2217008120}{Biological neurons act as generalization filters in reservoir computing}, Proceedings of the National Academy of Sciences 120~(25) (2023) e2217008120, publisher: Proceedings of the National Academy of Sciences.
\newblock \href {https://doi.org/10.1073/pnas.2217008120} {\path{doi:10.1073/pnas.2217008120}}.
\newline\urlprefix\url{https://www.pnas.org/doi/10.1073/pnas.2217008120}

\bibitem{hartung_baltimore_2023}
T.~Hartung, L.~Smirnova, I.~E. Morales~Pantoja, A.~Akwaboah, D.-M. Alam El~Din, C.~A. Berlinicke, J.~L. Boyd, B.~S. Caffo, B.~Cappiello, T.~Cohen-Karni, J.~L. Curley, R.~Etienne-Cummings, R.~Dastgheyb, D.~H. Gracias, F.~Gilbert, C.~W. Habela, F.~Han, T.~D. Harris, K.~Herrmann, E.~J. Hill, Q.~Huang, R.~E. Jabbour, E.~C. Johnson, B.~J. Kagan, C.~Krall, A.~Levchenko, P.~Locke, A.~Maertens, M.~Metea, A.~R. Muotri, R.~Parri, B.~L. Paulhamus, J.~D. Plotkin, P.~Roach, J.~C. Romero, J.~C. Schwamborn, F.~Sillé, A.~S. Szalay, K.~Tsaioun, D.~Tornero, J.~T. Vogelstein, K.~J. Wahlin, D.~J. Zack, \href{https://www.frontiersin.org/journals/science/articles/10.3389/fsci.2023.1068159/full}{The {Baltimore} declaration toward the exploration of organoid intelligence}, Frontiers in Science 1, publisher: Frontiers (Feb. 2023).
\newblock \href {https://doi.org/10.3389/fsci.2023.1068159} {\path{doi:10.3389/fsci.2023.1068159}}.
\newline\urlprefix\url{https://www.frontiersin.org/journals/science/articles/10.3389/fsci.2023.1068159/full}

\bibitem{morales_pantoja_first_2023}
I.~E. Morales~Pantoja, L.~Smirnova, A.~R. Muotri, K.~J. Wahlin, J.~Kahn, J.~L. Boyd, D.~H. Gracias, T.~D. Harris, T.~Cohen-Karni, B.~S. Caffo, A.~S. Szalay, F.~Han, D.~J. Zack, R.~Etienne-Cummings, A.~Akwaboah, J.~C. Romero, D.-M. Alam El~Din, J.~D. Plotkin, B.~L. Paulhamus, E.~C. Johnson, F.~Gilbert, J.~L. Curley, B.~Cappiello, J.~C. Schwamborn, E.~J. Hill, P.~Roach, D.~Tornero, C.~Krall, R.~Parri, F.~Sillé, A.~Levchenko, R.~E. Jabbour, B.~J. Kagan, C.~A. Berlinicke, Q.~Huang, A.~Maertens, K.~Herrmann, K.~Tsaioun, R.~Dastgheyb, C.~W. Habela, J.~T. Vogelstein, T.~Hartung, \href{https://www.frontiersin.org/journals/artificial-intelligence/articles/10.3389/frai.2023.1116870/full}{First {Organoid} {Intelligence} ({OI}) workshop to form an {OI} community}, Frontiers in Artificial Intelligence 6, publisher: Frontiers (Feb. 2023).
\newblock \href {https://doi.org/10.3389/frai.2023.1116870} {\path{doi:10.3389/frai.2023.1116870}}.
\newline\urlprefix\url{https://www.frontiersin.org/journals/artificial-intelligence/articles/10.3389/frai.2023.1116870/full}

\bibitem{koch_next-generation_2022}
C.~Koch, K.~Svoboda, A.~Bernard, M.~A. Basso, A.~K. Churchland, A.~L. Fairhall, P.~A. Groblewski, J.~A. Lecoq, Z.~F. Mainen, M.~W. Mathis, S.~R. Olsen, J.~W. Phillips, A.~Pouget, S.~Saxena, J.~H. Siegle, A.~M. Zador, Next-generation brain observatories, Neuron 110~(22) (2022) 3661--3666.
\newblock \href {https://doi.org/10.1016/j.neuron.2022.09.033} {\path{doi:10.1016/j.neuron.2022.09.033}}.

\bibitem{de_vries_sharing_2023}
S.~E. de~Vries, J.~H. Siegle, C.~Koch, \href{https://doi.org/10.7554/eLife.85550}{Sharing neurophysiology data from the {Allen} {Brain} {Observatory}}, eLife 12 (2023) e85550, publisher: eLife Sciences Publications, Ltd.
\newblock \href {https://doi.org/10.7554/eLife.85550} {\path{doi:10.7554/eLife.85550}}.
\newline\urlprefix\url{https://doi.org/10.7554/eLife.85550}

\bibitem{gillon_open_2024}
C.~J. Gillon, C.~Baker, R.~Ly, E.~Balzani, B.~W. Brunton, M.~Schottdorf, S.~Ghosh, N.~Dehghani, \href{https://www.ncbi.nlm.nih.gov/pmc/articles/PMC11247910/}{Open {Data} {In} {Neurophysiology}: {Advancements}, {Solutions} \& {Challenges}}, ArXiv (2024) arXiv:2407.00976v1.
\newline\urlprefix\url{https://www.ncbi.nlm.nih.gov/pmc/articles/PMC11247910/}

\bibitem{jordan_open_2024}
F.~D. Jordan, M.~Kutter, J.-M. Comby, F.~Brozzi, E.~Kurtys, \href{https://www.frontiersin.org/journals/artificial-intelligence/articles/10.3389/frai.2024.1376042/full}{Open and remotely accessible {Neuroplatform} for research in wetware computing}, Frontiers in Artificial Intelligence 7, publisher: Frontiers (May 2024).
\newblock \href {https://doi.org/10.3389/frai.2024.1376042} {\path{doi:10.3389/frai.2024.1376042}}.
\newline\urlprefix\url{https://www.frontiersin.org/journals/artificial-intelligence/articles/10.3389/frai.2024.1376042/full}

\bibitem{elliott_internet-connected_2023}
M.~A.~T. Elliott, H.~E. Schweiger, A.~Robbins, S.~Vera-Choqqueccota, D.~Ehrlich, S.~Hernandez, K.~Voitiuk, J.~Geng, J.~L. Sevetson, C.~Core, Y.~M. Rosen, M.~Teodorescu, N.~O. Wagner, D.~Haussler, M.~A. Mostajo-Radji, \href{https://www.eneuro.org/content/10/12/ENEURO.0308-23.2023}{Internet-{Connected} {Cortical} {Organoids} for {Project}-{Based} {Stem} {Cell} and {Neuroscience} {Education}}, eNeuro 10~(12), publisher: Society for Neuroscience Section: Research Article: New Research (Dec. 2023).
\newblock \href {https://doi.org/10.1523/ENEURO.0308-23.2023} {\path{doi:10.1523/ENEURO.0308-23.2023}}.
\newline\urlprefix\url{https://www.eneuro.org/content/10/12/ENEURO.0308-23.2023}

\bibitem{voitiuk_feedback-driven_2024}
K.~Voitiuk, S.~T. Seiler, M.~Pessoa~de Melo, J.~Geng, S.~Hernandez, H.~E. Schweiger, J.~L. Sevetson, D.~F. Parks, A.~Robbins, S.~Torres-Montoya, D.~Ehrlich, M.~A.~T. Elliott, T.~Sharf, D.~Haussler, M.~A. Mostajo-Radji, S.~R. Salama, M.~Teodorescu, A feedback-driven {IoT} microfluidic, electrophysiology, and imaging platform for brain organoid studies, bioRxiv: The Preprint Server for Biology (2024) 2024.03.15.585237\href {https://doi.org/10.1101/2024.03.15.585237} {\path{doi:10.1101/2024.03.15.585237}}.

\bibitem{zhang_mind_2024}
X.~Zhang, Z.~Dou, S.~H. Kim, G.~Upadhyay, D.~Havert, S.~Kang, K.~Kazemi, K.-Y. Huang, O.~Aydin, R.~Huang, S.~Rahman, A.~Ellis-Mohr, H.~A. Noblet, K.~H. Lim, H.~J. Chung, H.~J. Gritton, M.~T.~A. Saif, H.~J. Kong, J.~M. Beggs, M.~Gazzola, Mind {In} {Vitro} {Platforms}: {Versatile}, {Scalable}, {Robust}, and {Open} {Solutions} to {Interfacing} with {Living} {Neurons}, Advanced Science (Weinheim, Baden-Wurttemberg, Germany) 11~(11) (2024) e2306826.
\newblock \href {https://doi.org/10.1002/advs.202306826} {\path{doi:10.1002/advs.202306826}}.

\bibitem{chong_system_2023}
H.~W. Chong, B.~J. Kagan, A.~C. Kitchen, \href{https://patents.google.com/patent/US20230133430A1/en}{System and method for training in vitro neurons} (May 2023).
\newline\urlprefix\url{https://patents.google.com/patent/US20230133430A1/en}

\bibitem{bray2009wetware}
D.~Bray, Wetware: a computer in every living cell, Yale University Press, 2009.

\bibitem{mukherjee_establishment_2023}
S.~Mukherjee, P.~Malik, T.~K. Mukherjee, \href{https://doi.org/10.1007/978-981-19-1731-8_2-1}{Establishment of a {Cell} {Culture} {Laboratory}}, in: T.~K. Mukherjee, P.~Malik, S.~Mukherjee (Eds.), Practical {Approach} to {Mammalian} {Cell} and {Organ} {Culture}, Springer Nature, Singapore, 2023, pp. 43--82.
\newblock \href {https://doi.org/10.1007/978-981-19-1731-8_2-1} {\path{doi:10.1007/978-981-19-1731-8_2-1}}.
\newline\urlprefix\url{https://doi.org/10.1007/978-981-19-1731-8_2-1}

\bibitem{harrison_general_1997}
M.~A. Harrison, I.~F. Rae, \href{https://www.cambridge.org/core/books/general-techniques-of-cell-culture/7BC2BBE04CAD334D5D223D13333BE35D}{General {Techniques} of {Cell} {Culture}}, Handbooks in {Practical} {Animal} {Cell} {Biology}, Cambridge University Press, Cambridge, 1997.
\newblock \href {https://doi.org/10.1017/CBO9780511623226} {\path{doi:10.1017/CBO9780511623226}}.
\newline\urlprefix\url{https://www.cambridge.org/core/books/general-techniques-of-cell-culture/7BC2BBE04CAD334D5D223D13333BE35D}

\bibitem{geraghty2014guidelines}
R.~Geraghty, A.~Capes-Davis, J.~Davis, J.~Downward, R.~Freshney, I.~Knezevic, R.~Lovell-Badge, J.~Masters, J.~Meredith, G.~Stacey, et~al., Guidelines for the use of cell lines in biomedical research, British journal of cancer 111~(6) (2014) 1021--1046.

\bibitem{engber2011c57}
D.~Engber, The trouble with {Black-6}: A tiny alcoholic takes over the lab., \url{https://www.slate.com/articles/health_and_science/the_mouse_trap/2011/11/black_6_lab_mice_and_the_history_of_biomedical_research.html}, accessed: 2024-07-19 (2011).

\bibitem{wang_recent_2017}
Q.~Wang, M.~A. Timberlake, K.~Prall, Y.~Dwivedi, \href{https://www.ncbi.nlm.nih.gov/pmc/articles/PMC5605906/}{The {Recent} {Progress} in {Animal} {Models} of {Depression}}, Progress in neuro-psychopharmacology \& biological psychiatry 77 (2017) 99--109.
\newblock \href {https://doi.org/10.1016/j.pnpbp.2017.04.008} {\path{doi:10.1016/j.pnpbp.2017.04.008}}.
\newline\urlprefix\url{https://www.ncbi.nlm.nih.gov/pmc/articles/PMC5605906/}

\bibitem{pollen_establishing_2019}
A.~A. Pollen, A.~Bhaduri, M.~G. Andrews, T.~J. Nowakowski, O.~S. Meyerson, M.~A. Mostajo-Radji, E.~Di~Lullo, B.~Alvarado, M.~Bedolli, M.~L. Dougherty, I.~T. Fiddes, Z.~N. Kronenberg, J.~Shuga, A.~A. Leyrat, J.~A. West, M.~Bershteyn, C.~B. Lowe, B.~J. Pavlovic, S.~R. Salama, D.~Haussler, E.~E. Eichler, A.~R. Kriegstein, Establishing {Cerebral} {Organoids} as {Models} of {Human}-{Specific} {Brain} {Evolution}, Cell 176~(4) (2019) 743--756.e17.
\newblock \href {https://doi.org/10.1016/j.cell.2019.01.017} {\path{doi:10.1016/j.cell.2019.01.017}}.

\bibitem{faltin_morphological_1985}
J.~Faltin, Z.~Lodin, J.~Hartman, B.~Foucaud, G.~Gombos, M.~Sensenbrenner, \href{https://www.sciencedirect.com/science/article/pii/0736574885900024}{Morphological maturation and survival of chicken and rat embryonic neurons on different culture substrata}, International Journal of Developmental Neuroscience 3~(2) (1985) 111--121.
\newblock \href {https://doi.org/10.1016/0736-5748(85)90002-4} {\path{doi:10.1016/0736-5748(85)90002-4}}.
\newline\urlprefix\url{https://www.sciencedirect.com/science/article/pii/0736574885900024}

\bibitem{andrews_mtor_2020}
M.~G. Andrews, L.~Subramanian, A.~R. Kriegstein, {mTOR} signaling regulates the morphology and migration of outer radial glia in developing human cortex, eLife 9 (2020) e58737.
\newblock \href {https://doi.org/10.7554/eLife.58737} {\path{doi:10.7554/eLife.58737}}.

\bibitem{perillo_spontaneous_2023}
M.~Perillo, A.~Punzo, C.~Caliceti, C.~Sell, A.~Lorenzini, The spontaneous immortalization probability of mammalian cell culture strains, as their proliferative capacity, correlates with species body mass, not longevity, Biomedical Journal 46~(3) (2023) 100596.
\newblock \href {https://doi.org/10.1016/j.bj.2023.100596} {\path{doi:10.1016/j.bj.2023.100596}}.

\bibitem{voloshin_practical_2023}
N.~Voloshin, P.~Tyurin-Kuzmin, M.~Karagyaur, Z.~Akopyan, K.~Kulebyakin, Practical {Use} of {Immortalized} {Cells} in {Medicine}: {Current} {Advances} and {Future} {Perspectives}, International Journal of Molecular Sciences 24~(16) (2023) 12716.
\newblock \href {https://doi.org/10.3390/ijms241612716} {\path{doi:10.3390/ijms241612716}}.

\bibitem{calles_effects_2006}
K.~Calles, I.~Svensson, E.~Lindskog, L.~Häggström, Effects of conditioned medium factors and passage number on {Sf9} cell physiology and productivity, Biotechnology Progress 22~(2) (2006) 394--400.
\newblock \href {https://doi.org/10.1021/bp050297a} {\path{doi:10.1021/bp050297a}}.

\bibitem{wang_neurod4_2023}
H.~Wang, P.~Zhao, Y.~Zhang, Z.~Chen, H.~Bao, W.~Qian, J.~Wu, Z.~Xing, X.~Hu, K.~Jin, Q.~Zhuge, J.~Yang, \href{https://www.nature.com/articles/s41420-023-01595-8}{{NeuroD4} converts glioblastoma cells into neuron-like cells through the {SLC7A11}-{GSH}-{GPX4} antioxidant axis}, Cell Death Discovery 9~(1) (2023) 1--12, publisher: Nature Publishing Group.
\newblock \href {https://doi.org/10.1038/s41420-023-01595-8} {\path{doi:10.1038/s41420-023-01595-8}}.
\newline\urlprefix\url{https://www.nature.com/articles/s41420-023-01595-8}

\bibitem{zhang_rapid_2013}
Y.~Zhang, C.~Pak, Y.~Han, H.~Ahlenius, Z.~Zhang, S.~Chanda, S.~Marro, C.~Patzke, C.~Acuna, J.~Covy, W.~Xu, N.~Yang, T.~Danko, L.~Chen, M.~Wernig, T.~C. Südhof, \href{https://www.cell.com/neuron/abstract/S0896-6273(13)00449-2}{Rapid {Single}-{Step} {Induction} of {Functional} {Neurons} from {Human} {Pluripotent} {Stem} {Cells}}, Neuron 78~(5) (2013) 785--798, publisher: Elsevier.
\newblock \href {https://doi.org/10.1016/j.neuron.2013.05.029} {\path{doi:10.1016/j.neuron.2013.05.029}}.
\newline\urlprefix\url{https://www.cell.com/neuron/abstract/S0896-6273(13)00449-2}

\bibitem{hu_neural_2010}
B.-Y. Hu, J.~P. Weick, J.~Yu, L.-X. Ma, X.-Q. Zhang, J.~A. Thomson, S.-C. Zhang, Neural differentiation of human induced pluripotent stem cells follows developmental principles but with variable potency, Proceedings of the National Academy of Sciences of the United States of America 107~(9) (2010) 4335--4340.
\newblock \href {https://doi.org/10.1073/pnas.0910012107} {\path{doi:10.1073/pnas.0910012107}}.

\bibitem{song_crosstalk_2018}
I.~Song, A.~Dityatev, Crosstalk between glia, extracellular matrix and neurons, Brain Research Bulletin 136 (2018) 101--108.
\newblock \href {https://doi.org/10.1016/j.brainresbull.2017.03.003} {\path{doi:10.1016/j.brainresbull.2017.03.003}}.

\bibitem{liu_coating_2020}
D.~Liu, N.~Pavathuparambil Abdul~Manaph, M.~Al-Hawwas, L.~Bobrovskaya, L.-L. Xiong, X.-F. Zhou, Coating {Materials} for {Neural} {Stem}/{Progenitor} {Cell} {Culture} and {Differentiation}, Stem Cells and Development 29~(8) (2020) 463--474.
\newblock \href {https://doi.org/10.1089/scd.2019.0288} {\path{doi:10.1089/scd.2019.0288}}.

\bibitem{egert_heart_2005}
U.~Egert, T.~Meyer, \href{https://doi.org/10.1007/3-540-26574-0_22}{Heart on a {Chip} — {Extracellular} {Multielectrode} {Recordings} from {Cardiac} {Myocytes} in {Vitro}}, in: S.~Dhein, F.~W. Mohr, M.~Delmar (Eds.), Practical {Methods} in {Cardiovascular} {Research}, Springer, Berlin, Heidelberg, 2005, pp. 432--453.
\newblock \href {https://doi.org/10.1007/3-540-26574-0_22} {\path{doi:10.1007/3-540-26574-0_22}}.
\newline\urlprefix\url{https://doi.org/10.1007/3-540-26574-0_22}

\bibitem{chad_cellular_1991}
J.~Chad, H.~Wheal (Eds.), Cellular {Neurobiology}: {A} {Practical} {Approach}, 0th Edition, IRL Press, Oxford, 1991.

\bibitem{duval_modeling_2017}
K.~Duval, H.~Grover, L.-H. Han, Y.~Mou, A.~F. Pegoraro, J.~Fredberg, Z.~Chen, Modeling {Physiological} {Events} in {2D} vs. {3D} {Cell} {Culture}, Physiology (Bethesda, Md.) 32~(4) (2017) 266--277.
\newblock \href {https://doi.org/10.1152/physiol.00036.2016} {\path{doi:10.1152/physiol.00036.2016}}.

\bibitem{verma_animal_2020}
A.~Verma, M.~Verma, A.~Singh, \href{https://www.ncbi.nlm.nih.gov/pmc/articles/PMC7325846/}{Animal tissue culture principles and applications}, Animal Biotechnology (2020) 269--293\href {https://doi.org/10.1016/B978-0-12-811710-1.00012-4} {\path{doi:10.1016/B978-0-12-811710-1.00012-4}}.
\newline\urlprefix\url{https://www.ncbi.nlm.nih.gov/pmc/articles/PMC7325846/}

\bibitem{laplaca2010three}
M.~C. LaPlaca, V.~N. Vernekar, J.~T. Shoemaker, D.~K. Cullen, W.~Coulter, Three-dimensional neuronal cultures, Methods in bioengineering: 3D tissue engineering (2010) 187--204.

\bibitem{lancaster_organogenesis_2014}
M.~A. Lancaster, J.~A. Knoblich, \href{https://www.science.org/doi/10.1126/science.1247125}{Organogenesis in a dish: {Modeling} development and disease using organoid technologies}, Science 345~(6194) (2014) 1247125, publisher: American Association for the Advancement of Science.
\newblock \href {https://doi.org/10.1126/science.1247125} {\path{doi:10.1126/science.1247125}}.
\newline\urlprefix\url{https://www.science.org/doi/10.1126/science.1247125}

\bibitem{matsui_vascularization_2021}
T.~K. Matsui, Y.~Tsuru, K.~Hasegawa, K.-I. Kuwako, Vascularization of human brain organoids, Stem Cells (Dayton, Ohio) 39~(8) (2021) 1017--1024.
\newblock \href {https://doi.org/10.1002/stem.3368} {\path{doi:10.1002/stem.3368}}.

\bibitem{zhang_vascularized_2021}
S.~Zhang, Z.~Wan, R.~D. Kamm, \href{https://pubs.rsc.org/en/content/articlelanding/2021/lc/d0lc01186j}{Vascularized organoids on a chip: strategies for engineering organoids with functional vasculature}, Lab on a Chip 21~(3) (2021) 473--488, publisher: The Royal Society of Chemistry.
\newblock \href {https://doi.org/10.1039/D0LC01186J} {\path{doi:10.1039/D0LC01186J}}.
\newline\urlprefix\url{https://pubs.rsc.org/en/content/articlelanding/2021/lc/d0lc01186j}

\bibitem{aras_assessment_2008}
M.~A. Aras, K.~A. Hartnett, E.~Aizenman, Assessment of cell viability in primary neuronal cultures, Current Protocols in Neuroscience Chapter 7 (2008) Unit 7.18.
\newblock \href {https://doi.org/10.1002/0471142301.ns0718s44} {\path{doi:10.1002/0471142301.ns0718s44}}.

\bibitem{pathak_quantum_2006}
S.~Pathak, E.~Cao, M.~C. Davidson, S.~Jin, G.~A. Silva, \href{https://www.ncbi.nlm.nih.gov/pmc/articles/PMC6674918/}{Quantum {Dot} {Applications} to {Neuroscience}: {New} {Tools} for {Probing} {Neurons} and {Glia}}, The Journal of Neuroscience 26~(7) (2006) 1893--1895.
\newblock \href {https://doi.org/10.1523/JNEUROSCI.3847-05.2006} {\path{doi:10.1523/JNEUROSCI.3847-05.2006}}.
\newline\urlprefix\url{https://www.ncbi.nlm.nih.gov/pmc/articles/PMC6674918/}

\bibitem{langdon_cell_2004}
S.~P. Langdon, Cell culture contamination: an overview, Methods in Molecular Medicine 88 (2004) 309--317.
\newblock \href {https://doi.org/10.1385/1-59259-406-9:309} {\path{doi:10.1385/1-59259-406-9:309}}.

\bibitem{nelson-rees_responsibility_2001}
W.~A. Nelson-Rees, \href{https://www.ncbi.nlm.nih.gov/pmc/articles/PMC1088478/}{Responsibility for truth in research.}, Philosophical Transactions of the Royal Society of London. Series B 356~(1410) (2001) 849--851.
\newblock \href {https://doi.org/10.1098/rstb.2001.0873} {\path{doi:10.1098/rstb.2001.0873}}.
\newline\urlprefix\url{https://www.ncbi.nlm.nih.gov/pmc/articles/PMC1088478/}

\bibitem{nam_vitro_2011}
Y.~Nam, B.~C. Wheeler, In vitro microelectrode array technology and neural recordings, Critical Reviews in Biomedical Engineering 39~(1) (2011) 45--61.
\newblock \href {https://doi.org/10.1615/critrevbiomedeng.v39.i1.40} {\path{doi:10.1615/critrevbiomedeng.v39.i1.40}}.

\bibitem{cameron_prolonged_2017}
M.~A. Cameron, O.~Kekesi, J.~W. Morley, A.~Bellot-Saez, S.~Kueh, P.~Breen, A.~van Schaik, J.~Tapson, Y.~Buskila, \href{https://www.ncbi.nlm.nih.gov/pmc/articles/PMC5408972/}{Prolonged {Incubation} of {Acute} {Neuronal} {Tissue} for {Electrophysiology} and {Calcium}-imaging}, Journal of Visualized Experiments : JoVE~(120) (2017) 55396.
\newblock \href {https://doi.org/10.3791/55396} {\path{doi:10.3791/55396}}.
\newline\urlprefix\url{https://www.ncbi.nlm.nih.gov/pmc/articles/PMC5408972/}

\bibitem{grodem_updated_2023}
S.~Grødem, I.~Nymoen, G.~H. Vatne, F.~S. Rogge, V.~Björnsdóttir, K.~K. Lensjø, M.~Fyhn, \href{https://www.nature.com/articles/s41467-023-36324-3}{An updated suite of viral vectors for in vivo calcium imaging using intracerebral and retro-orbital injections in male mice}, Nature Communications 14~(1) (2023) 608, publisher: Nature Publishing Group.
\newblock \href {https://doi.org/10.1038/s41467-023-36324-3} {\path{doi:10.1038/s41467-023-36324-3}}.
\newline\urlprefix\url{https://www.nature.com/articles/s41467-023-36324-3}

\bibitem{antic_voltage_2016}
S.~D. Antic, R.~M. Empson, T.~Knöpfel, Voltage imaging to understand connections and functions of neuronal circuits, Journal of Neurophysiology 116~(1) (2016) 135--152.
\newblock \href {https://doi.org/10.1152/jn.00226.2016} {\path{doi:10.1152/jn.00226.2016}}.

\bibitem{adam_voltage_2019}
Y.~Adam, J.~J. Kim, S.~Lou, Y.~Zhao, M.~E. Xie, D.~Brinks, H.~Wu, M.~A. Mostajo-Radji, S.~Kheifets, V.~Parot, S.~Chettih, K.~J. Williams, B.~Gmeiner, S.~L. Farhi, L.~Madisen, E.~K. Buchanan, I.~Kinsella, D.~Zhou, L.~Paninski, C.~D. Harvey, H.~Zeng, P.~Arlotta, R.~E. Campbell, A.~E. Cohen, \href{https://www.nature.com/articles/s41586-019-1166-7}{Voltage imaging and optogenetics reveal behaviour-dependent changes in hippocampal dynamics}, Nature 569~(7756) (2019) 413--417, publisher: Nature Publishing Group.
\newblock \href {https://doi.org/10.1038/s41586-019-1166-7} {\path{doi:10.1038/s41586-019-1166-7}}.
\newline\urlprefix\url{https://www.nature.com/articles/s41586-019-1166-7}

\bibitem{jun_fully_2017}
J.~J. Jun, N.~A. Steinmetz, J.~H. Siegle, D.~J. Denman, M.~Bauza, B.~Barbarits, A.~K. Lee, C.~A. Anastassiou, A.~Andrei, Ã.~Aydın, M.~Barbic, T.~J. Blanche, V.~Bonin, J.~Couto, B.~Dutta, S.~L. Gratiy, D.~A. Gutnisky, M.~Häusser, B.~Karsh, P.~Ledochowitsch, C.~M. Lopez, C.~Mitelut, S.~Musa, M.~Okun, M.~Pachitariu, J.~Putzeys, P.~D. Rich, C.~Rossant, W.-l. Sun, K.~Svoboda, M.~Carandini, K.~D. Harris, C.~Koch, J.~O'Keefe, T.~D. Harris, \href{https://www.nature.com/articles/nature24636}{Fully integrated silicon probes for high-density recording of neural activity}, Nature 551~(7679) (2017) 232--236, publisher: Nature Publishing Group.
\newblock \href {https://doi.org/10.1038/nature24636} {\path{doi:10.1038/nature24636}}.
\newline\urlprefix\url{https://www.nature.com/articles/nature24636}

\bibitem{wagenaar_extremely_2006}
D.~A. Wagenaar, J.~Pine, S.~M. Potter, \href{https://doi.org/10.1186/1471-2202-7-11}{An extremely rich repertoire of bursting patterns during the development of cortical cultures}, BMC Neuroscience 7~(1) (2006) 11.
\newblock \href {https://doi.org/10.1186/1471-2202-7-11} {\path{doi:10.1186/1471-2202-7-11}}.
\newline\urlprefix\url{https://doi.org/10.1186/1471-2202-7-11}

\bibitem{muller_sub-millisecond_2012}
J.~Müller, D.~J. Bakkum, A.~Hierlemann, Sub-millisecond closed-loop feedback stimulation between arbitrary sets of individual neurons, Frontiers in Neural Circuits 6 (2012) 121.
\newblock \href {https://doi.org/10.3389/fncir.2012.00121} {\path{doi:10.3389/fncir.2012.00121}}.

\bibitem{newman_closed-loop_2012}
J.~P. Newman, R.~Zeller-Townson, M.-F. Fong, S.~Arcot~Desai, R.~E. Gross, S.~M. Potter, Closed-{Loop}, {Multichannel} {Experimentation} {Using} the {Open}-{Source} {NeuroRighter} {Electrophysiology} {Platform}, Frontiers in Neural Circuits 6 (2012) 98.
\newblock \href {https://doi.org/10.3389/fncir.2012.00098} {\path{doi:10.3389/fncir.2012.00098}}.

\bibitem{corticallabs1}
Meet the 'man with 20 brains' powering this game-changing aussie start-up, \url{https://www.forbes.com.au/news/innovation/meet-the-man-with-20-brains-powering-this-game-changing-aussie-start-up/}, accessed: 2024-07-19 (2023).

\bibitem{li_organoid_2020}
Y.~Li, P.~Tang, S.~Cai, J.~Peng, G.~Hua, Organoid based personalized medicine: from bench to bedside, Cell Regeneration (London, England) 9~(1) (2020) 21.
\newblock \href {https://doi.org/10.1186/s13619-020-00059-z} {\path{doi:10.1186/s13619-020-00059-z}}.

\bibitem{haring_microphysiological_2017}
A.~P. Haring, H.~Sontheimer, B.~N. Johnson, Microphysiological {Human} {Brain} and {Neural} {Systems}-on-a-{Chip}: {Potential} {Alternatives} to {Small} {Animal} {Models} and {Emerging} {Platforms} for {Drug} {Discovery} and {Personalized} {Medicine}, Stem Cell Reviews and Reports 13~(3) (2017) 381--406.
\newblock \href {https://doi.org/10.1007/s12015-017-9738-0} {\path{doi:10.1007/s12015-017-9738-0}}.

\bibitem{kagan_scientific_2023}
B.~J. Kagan, A.~Razi, A.~Bhat, A.~C. Kitchen, N.~T. Tran, F.~Habibollahi, M.~Khajehnejad, B.~J. Parker, B.~Rollo, K.~J. Friston, Scientific communication and the semantics of sentience, Neuron 111~(5) (2023) 606--607.
\newblock \href {https://doi.org/10.1016/j.neuron.2023.02.008} {\path{doi:10.1016/j.neuron.2023.02.008}}.

\bibitem{rommelfanger_conceptual_2023}
K.~S. Rommelfanger, K.~M. Ramos, A.~Salles, Conceptual conundrums for neuroscience, Neuron 111~(5) (2023) 608--609.
\newblock \href {https://doi.org/10.1016/j.neuron.2023.02.016} {\path{doi:10.1016/j.neuron.2023.02.016}}.

\bibitem{pereira_neural_2023}
A.~Pereira, J.~W. Garcia, A.~Muotri, \href{https://www.mdpi.com/2673-4087/4/1/4}{Neural {Stimulation} of {Brain} {Organoids} with {Dynamic} {Patterns}: {A} {Sentiomics} {Approach} {Directed} to {Regenerative} {Neuromedicine}}, NeuroSci 4~(1) (2023) 31--42, number: 1 Publisher: Multidisciplinary Digital Publishing Institute.
\newblock \href {https://doi.org/10.3390/neurosci4010004} {\path{doi:10.3390/neurosci4010004}}.
\newline\urlprefix\url{https://www.mdpi.com/2673-4087/4/1/4}

\bibitem{kagan_toward_2024}
B.~J. Kagan, M.~Mahlis, A.~Bhat, J.~Bongard, V.~M. Cole, P.~Corlett, C.~Gyngell, T.~Hartung, B.~Jupp, M.~Levin, T.~Lysaght, N.~Opie, A.~Razi, L.~Smirnova, I.~Tennant, P.~T. Wade, G.~Wang, \href{https://www.sciencedirect.com/science/article/pii/S2666675824000961}{Toward a nomenclature consensus for diverse intelligent systems: {Call} for collaboration}, The Innovation 5~(5) (2024) 100658.
\newblock \href {https://doi.org/10.1016/j.xinn.2024.100658} {\path{doi:10.1016/j.xinn.2024.100658}}.
\newline\urlprefix\url{https://www.sciencedirect.com/science/article/pii/S2666675824000961}

\bibitem{kagan_embodied_2024}
B.~J. Kagan, A.~Loeffler, J.~L. Boyd, J.~Savulescu, \href{https://www.jneurosci.org/content/44/15/e0431242024}{Embodied {Neural} {Systems} {Can} {Enable} {Iterative} {Investigations} of {Morally} {Relevant} {States}}, Journal of Neuroscience 44~(15), publisher: Society for Neuroscience Section: Commentary (Apr. 2024).
\newblock \href {https://doi.org/10.1523/JNEUROSCI.0431-24.2024} {\path{doi:10.1523/JNEUROSCI.0431-24.2024}}.
\newline\urlprefix\url{https://www.jneurosci.org/content/44/15/e0431242024}

\bibitem{o2006setting}
R.~O'Connor, L.~O'Driscoll, Setting up a cell culture laboratory, in: Cell Biology, Elsevier, 2006, pp. 5--11.

\bibitem{richmond2009biosafety}
J.~Y. Richmond, R.~W. McKinney, Biosafety in microbiological and biomedical laboratories, US Government Printing Office, 2009.

\bibitem{goodman2007centrifuge}
T.~Goodman, Centrifuge rotor selection and maintenance, American laboratory 39~(12) (2007) 12.

\bibitem{petrov_effect_2018}
A.~I. Petrov, M.~V. Razuvaeva, \href{https://doi.org/10.1134/S1063784218100183}{Effect of {Temperature} on {Metabolism} and {Lifespan} in {Several} {Homeothermic} {Animals}}, Technical Physics 63~(10) (2018) 1410--1414.
\newblock \href {https://doi.org/10.1134/S1063784218100183} {\path{doi:10.1134/S1063784218100183}}.
\newline\urlprefix\url{https://doi.org/10.1134/S1063784218100183}

\bibitem{geneva_normal_2019}
I.~I. Geneva, B.~Cuzzo, T.~Fazili, W.~Javaid, Normal {Body} {Temperature}: {A} {Systematic} {Review}, Open Forum Infectious Diseases 6~(4) (2019) ofz032.
\newblock \href {https://doi.org/10.1093/ofid/ofz032} {\path{doi:10.1093/ofid/ofz032}}.

\bibitem{ukhealth2020co2conc}
{UK Health Security Agency}, {CO\textsubscript{2} concentration and pH control in the cell culture laboratory}, \url{https://www.culturecollections.org.uk/culture-collection-news/co2-concentration-and-ph-control-in-the-cell-culture-laboratory/}, accessed: July 19, 2024.

\bibitem{mestres_factors_2021}
E.~Mestres, M.~García-Jiménez, A.~Casals, J.~Cohen, M.~Acacio, A.~Villamar, Q.~Matia-Algué, G.~Calderón, N.~Costa-Borges, \href{https://doi.org/10.1093/humrep/deaa370}{Factors of the human embryo culture system that may affect media evaporation and osmolality}, Human Reproduction 36~(3) (2021) 605--613.
\newblock \href {https://doi.org/10.1093/humrep/deaa370} {\path{doi:10.1093/humrep/deaa370}}.
\newline\urlprefix\url{https://doi.org/10.1093/humrep/deaa370}

\bibitem{ibidi2022humidity}
Application note 12: Avoiding evaporation: Humidity control in cell culture, \url{https://ibidi.com/img/cms/support/AN/AN12_Avoiding_evaporation.pdf}, accessed: 2024-07-19 (2022).

\bibitem{chakrabarti_defects_2007}
L.~Chakrabarti, Z.~Galdzicki, T.~F. Haydar, Defects in embryonic neurogenesis and initial synapse formation in the forebrain of the {Ts65Dn} mouse model of {Down} syndrome, The Journal of Neuroscience: The Official Journal of the Society for Neuroscience 27~(43) (2007) 11483--11495.
\newblock \href {https://doi.org/10.1523/JNEUROSCI.3406-07.2007} {\path{doi:10.1523/JNEUROSCI.3406-07.2007}}.

\bibitem{dumanis_apoe_2011}
S.~B. Dumanis, H.-J. Cha, J.~M. Song, J.~H. Trotter, M.~Spitzer, J.-Y. Lee, E.~J. Weeber, R.~S. Turner, D.~T.~S. Pak, G.~W. Rebeck, H.-S. Hoe, {ApoE} receptor 2 regulates synapse and dendritic spine formation, PloS One 6~(2) (2011) e17203.
\newblock \href {https://doi.org/10.1371/journal.pone.0017203} {\path{doi:10.1371/journal.pone.0017203}}.

\bibitem{kandel_principles_2012}
E.~R. Kandel, J.~H. Schwartz, T.~M. Jessell, S.~A. Siegelbaum, A.~J. Hudspeth (Eds.), Principles of {Neural} {Science}, {Fifth} {Edition}, 5th Edition, McGraw-Hill Education / Medical, New York, 2012.

\bibitem{gilbert_chapter_2018}
T.~L. Gilbert, L.~Ng, \href{https://www.sciencedirect.com/science/article/pii/B9780128040782000039}{Chapter 3 - {The} {Allen} {Brain} {Atlas}: {Toward} {Understanding} {Brain} {Behavior} and {Function} {Through} {Data} {Acquisition}, {Visualization}, {Analysis}, and {Integration}}, in: R.~T. Gerlai (Ed.), Molecular-{Genetic} and {Statistical} {Techniques} for {Behavioral} and {Neural} {Research}, Academic Press, San Diego, 2018, pp. 51--72.
\newblock \href {https://doi.org/10.1016/B978-0-12-804078-2.00003-9} {\path{doi:10.1016/B978-0-12-804078-2.00003-9}}.
\newline\urlprefix\url{https://www.sciencedirect.com/science/article/pii/B9780128040782000039}

\bibitem{thompson_high-resolution_2014}
C.~L. Thompson, L.~Ng, V.~Menon, S.~Martinez, C.-K. Lee, K.~Glattfelder, S.~M. Sunkin, A.~Henry, C.~Lau, C.~Dang, R.~Garcia-Lopez, A.~Martinez-Ferre, A.~Pombero, J.~L.~R. Rubenstein, W.~B. Wakeman, J.~Hohmann, N.~Dee, A.~J. Sodt, R.~Young, K.~Smith, T.-N. Nguyen, J.~Kidney, L.~Kuan, A.~Jeromin, A.~Kaykas, J.~Miller, D.~Page, G.~Orta, A.~Bernard, Z.~Riley, S.~Smith, P.~Wohnoutka, M.~J. Hawrylycz, L.~Puelles, A.~R. Jones, A high-resolution spatiotemporal atlas of gene expression of the developing mouse brain, Neuron 83~(2) (2014) 309--323.
\newblock \href {https://doi.org/10.1016/j.neuron.2014.05.033} {\path{doi:10.1016/j.neuron.2014.05.033}}.

\bibitem{kamioka_spontaneous_1996}
H.~Kamioka, E.~Maeda, Y.~Jimbo, H.~P. Robinson, A.~Kawana, Spontaneous periodic synchronized bursting during formation of mature patterns of connections in cortical cultures, Neuroscience Letters 206~(2-3) (1996) 109--112.
\newblock \href {https://doi.org/10.1016/s0304-3940(96)12448-4} {\path{doi:10.1016/s0304-3940(96)12448-4}}.

\bibitem{curzer_three_2016}
H.~J. Curzer, G.~Perry, M.~C. Wallace, D.~Perry, The {Three} {Rs} of {Animal} {Research}: {What} they {Mean} for the {Institutional} {Animal} {Care} and {Use} {Committee} and {Why}, Science and Engineering Ethics 22~(2) (2016) 549--565.
\newblock \href {https://doi.org/10.1007/s11948-015-9659-8} {\path{doi:10.1007/s11948-015-9659-8}}.

\bibitem{bhaduri_outer_2020}
A.~Bhaduri, E.~Di~Lullo, D.~Jung, S.~Müller, E.~E. Crouch, C.~S. Espinosa, T.~Ozawa, B.~Alvarado, J.~Spatazza, C.~R. Cadwell, G.~Wilkins, D.~Velmeshev, S.~J. Liu, M.~Malatesta, M.~G. Andrews, M.~A. Mostajo-Radji, E.~J. Huang, T.~J. Nowakowski, D.~A. Lim, A.~Diaz, D.~R. Raleigh, A.~R. Kriegstein, Outer {Radial} {Glia}-like {Cancer} {Stem} {Cells} {Contribute} to {Heterogeneity} of {Glioblastoma}, Cell Stem Cell 26~(1) (2020) 48--63.e6.
\newblock \href {https://doi.org/10.1016/j.stem.2019.11.015} {\path{doi:10.1016/j.stem.2019.11.015}}.

\bibitem{kumar_data_2018}
M.~Kumar, A.~Katyal, Data on retinoic acid and reduced serum concentration induced differentiation of {Neuro}-2a neuroblastoma cells, Data in Brief 21 (2018) 2435--2440.
\newblock \href {https://doi.org/10.1016/j.dib.2018.11.097} {\path{doi:10.1016/j.dib.2018.11.097}}.

\bibitem{velasco_individual_2019}
S.~Velasco, A.~J. Kedaigle, S.~K. Simmons, A.~Nash, M.~Rocha, G.~Quadrato, B.~Paulsen, L.~Nguyen, X.~Adiconis, A.~Regev, J.~Z. Levin, P.~Arlotta, \href{https://www.nature.com/articles/s41586-019-1289-x}{Individual brain organoids reproducibly form cell diversity of the human cerebral cortex}, Nature 570~(7762) (2019) 523--527, publisher: Nature Publishing Group.
\newblock \href {https://doi.org/10.1038/s41586-019-1289-x} {\path{doi:10.1038/s41586-019-1289-x}}.
\newline\urlprefix\url{https://www.nature.com/articles/s41586-019-1289-x}

\bibitem{nowakowski_cerebral_2022}
T.~J. Nowakowski, S.~R. Salama, Cerebral {Organoids} as an {Experimental} {Platform} for {Human} {Neurogenomics}, Cells 11~(18) (2022) 2803.
\newblock \href {https://doi.org/10.3390/cells11182803} {\path{doi:10.3390/cells11182803}}.

\bibitem{amin_generating_2023}
N.~D. Amin, K.~W. Kelley, J.~Hao, Y.~Miura, G.~Narazaki, T.~Li, P.~McQueen, S.~Kulkarni, S.~Pavlov, S.~P. Paşca, Generating human neural diversity with a multiplexed morphogen screen in organoids, bioRxiv: The Preprint Server for Biology (2023) 2023.05.31.541819\href {https://doi.org/10.1101/2023.05.31.541819} {\path{doi:10.1101/2023.05.31.541819}}.

\bibitem{zettler_establishing_2007}
P.~Zettler, L.~E. Wolf, B.~Lo, Establishing procedures for institutional oversight of stem cell research, Academic Medicine: Journal of the Association of American Medical Colleges 82~(1) (2007) 6--10.
\newblock \href {https://doi.org/10.1097/01.ACM.0000250025.17863.bf} {\path{doi:10.1097/01.ACM.0000250025.17863.bf}}.

\bibitem{yoshiki_sasai_self-organization_nodate}
T.~K. Yoshiki~Sasai, \href{https://www.pnas.org/doi/10.1073/pnas.1315710110}{Self-organization of axial polarity, inside-out layer pattern, and species-specific progenitor dynamics in human {ES} cell–derived neocortex}, iSBN: 9781315710112.
\newline\urlprefix\url{https://www.pnas.org/doi/10.1073/pnas.1315710110}

\bibitem{eiraku_self-organized_2008}
M.~Eiraku, K.~Watanabe, M.~Matsuo-Takasaki, M.~Kawada, S.~Yonemura, M.~Matsumura, T.~Wataya, A.~Nishiyama, K.~Muguruma, Y.~Sasai, Self-organized formation of polarized cortical tissues from {ESCs} and its active manipulation by extrinsic signals, Cell Stem Cell 3~(5) (2008) 519--532.
\newblock \href {https://doi.org/10.1016/j.stem.2008.09.002} {\path{doi:10.1016/j.stem.2008.09.002}}.

\bibitem{sasai_cytosystems_2013}
Y.~Sasai, \href{https://www.nature.com/articles/nature11859}{Cytosystems dynamics in self-organization of tissue architecture}, Nature 493~(7432) (2013) 318--326, publisher: Nature Publishing Group.
\newblock \href {https://doi.org/10.1038/nature11859} {\path{doi:10.1038/nature11859}}.
\newline\urlprefix\url{https://www.nature.com/articles/nature11859}

\bibitem{brewer_nbactiv4_2008}
G.~J. Brewer, T.~T. Jones, M.~D. Boehler, B.~C. Wheeler, \href{https://www.ncbi.nlm.nih.gov/pmc/articles/PMC2393548/}{{NbActiv4} medium improvement to {Neurobasal}/{B27} increases neuron synapse densities and network spike rates on multielectrode arrays}, Journal of neuroscience methods 170~(2) (2008) 181--187.
\newblock \href {https://doi.org/10.1016/j.jneumeth.2008.01.009} {\path{doi:10.1016/j.jneumeth.2008.01.009}}.
\newline\urlprefix\url{https://www.ncbi.nlm.nih.gov/pmc/articles/PMC2393548/}

\bibitem{bardy_neuronal_2015}
C.~Bardy, M.~van~den Hurk, T.~Eames, C.~Marchand, R.~V. Hernandez, M.~Kellogg, M.~Gorris, B.~Galet, V.~Palomares, J.~Brown, A.~G. Bang, J.~Mertens, L.~Böhnke, L.~Boyer, S.~Simon, F.~H. Gage, \href{https://www.pnas.org/doi/10.1073/pnas.1504393112}{Neuronal medium that supports basic synaptic functions and activity of human neurons in vitro}, Proceedings of the National Academy of Sciences 112~(20) (2015) E2725--E2734, publisher: Proceedings of the National Academy of Sciences.
\newblock \href {https://doi.org/10.1073/pnas.1504393112} {\path{doi:10.1073/pnas.1504393112}}.
\newline\urlprefix\url{https://www.pnas.org/doi/10.1073/pnas.1504393112}

\bibitem{ghosh_requirement_1994}
A.~Ghosh, J.~Carnahan, M.~E. Greenberg, Requirement for {BDNF} in activity-dependent survival of cortical neurons, Science (New York, N.Y.) 263~(5153) (1994) 1618--1623.
\newblock \href {https://doi.org/10.1126/science.7907431} {\path{doi:10.1126/science.7907431}}.

\bibitem{park_modulation_2024}
Y.~Park, S.~Hernandez, C.~O. Hernandez, H.~E. Schweiger, H.~Li, K.~Voitiuk, H.~Dechiraju, N.~Hawthorne, E.~M. Muzzy, J.~A. Selberg, F.~N. Sullivan, R.~Urcuyo, S.~R. Salama, E.~Aslankoohi, H.~J. Knight, M.~Teodorescu, M.~A. Mostajo-Radji, M.~Rolandi, \href{https://www.cell.com/cell-reports-methods/abstract/S2667-2375(23)00372-7}{Modulation of neuronal activity in cortical organoids with bioelectronic delivery of ions and neurotransmitters}, Cell Reports Methods 4~(1), publisher: Elsevier (Jan. 2024).
\newblock \href {https://doi.org/10.1016/j.crmeth.2023.100686} {\path{doi:10.1016/j.crmeth.2023.100686}}.
\newline\urlprefix\url{https://www.cell.com/cell-reports-methods/abstract/S2667-2375(23)00372-7}

\bibitem{quadrato_cell_2017}
G.~Quadrato, T.~Nguyen, E.~Z. Macosko, J.~L. Sherwood, S.~Min~Yang, D.~R. Berger, N.~Maria, J.~Scholvin, M.~Goldman, J.~P. Kinney, E.~S. Boyden, J.~W. Lichtman, Z.~M. Williams, S.~A. McCarroll, P.~Arlotta, \href{https://www.nature.com/articles/nature22047}{Cell diversity and network dynamics in photosensitive human brain organoids}, Nature 545~(7652) (2017) 48--53, publisher: Nature Publishing Group.
\newblock \href {https://doi.org/10.1038/nature22047} {\path{doi:10.1038/nature22047}}.
\newline\urlprefix\url{https://www.nature.com/articles/nature22047}

\bibitem{petasch_low-pressure_1997}
W.~Petasch, B.~Kegel, H.~Schmid, K.~Lendenmann, H.~U. Keller, \href{https://www.sciencedirect.com/science/article/pii/S0257897297001436}{Low-pressure plasma cleaning: a process for precision cleaning applications}, Surface and Coatings Technology 97~(1) (1997) 176--181.
\newblock \href {https://doi.org/10.1016/S0257-8972(97)00143-6} {\path{doi:10.1016/S0257-8972(97)00143-6}}.
\newline\urlprefix\url{https://www.sciencedirect.com/science/article/pii/S0257897297001436}

\bibitem{brewer_optimized_1993}
G.~J. Brewer, J.~R. Torricelli, E.~K. Evege, P.~J. Price, Optimized survival of hippocampal neurons in {B27}-supplemented {Neurobasal}, a new serum-free medium combination, Journal of Neuroscience Research 35~(5) (1993) 567--576.
\newblock \href {https://doi.org/10.1002/jnr.490350513} {\path{doi:10.1002/jnr.490350513}}.

\bibitem{drexler_mycoplasma_2002}
H.~G. Drexler, C.~C. Uphoff, Mycoplasma contamination of cell cultures: {Incidence}, sources, effects, detection, elimination, prevention, Cytotechnology 39~(2) (2002) 75--90.
\newblock \href {https://doi.org/10.1023/A:1022913015916} {\path{doi:10.1023/A:1022913015916}}.

\bibitem{gao_nanotechnology_2021}
J.~Gao, C.~Liao, S.~Liu, T.~Xia, G.~Jiang, Nanotechnology: new opportunities for the development of patch-clamps, Journal of Nanobiotechnology 19~(1) (2021) 97.
\newblock \href {https://doi.org/10.1186/s12951-021-00841-4} {\path{doi:10.1186/s12951-021-00841-4}}.

\bibitem{hong_novel_2019}
G.~Hong, C.~M. Lieber, \href{https://www.nature.com/articles/s41583-019-0140-6}{Novel electrode technologies for neural recordings}, Nature Reviews Neuroscience 20~(6) (2019) 330--345, publisher: Nature Publishing Group.
\newblock \href {https://doi.org/10.1038/s41583-019-0140-6} {\path{doi:10.1038/s41583-019-0140-6}}.
\newline\urlprefix\url{https://www.nature.com/articles/s41583-019-0140-6}

\bibitem{sileo_electrical_2013}
L.~Sileo, F.~Pisanello, L.~Quarta, A.~Maccione, A.~Simi, L.~Berdondini, M.~De~Vittorio, L.~Martiradonna, \href{https://www.sciencedirect.com/science/article/pii/S016793171300381X}{Electrical coupling of mammalian neurons to microelectrodes with {3D} nanoprotrusions}, Microelectronic Engineering 111 (2013) 384--390.
\newblock \href {https://doi.org/10.1016/j.mee.2013.03.152} {\path{doi:10.1016/j.mee.2013.03.152}}.
\newline\urlprefix\url{https://www.sciencedirect.com/science/article/pii/S016793171300381X}

\bibitem{steins_flexible_2022}
H.~Steins, M.~Mierzejewski, L.~Brauns, A.~Stumpf, A.~Kohler, G.~Heusel, A.~Corna, T.~Herrmann, P.~D. Jones, G.~Zeck, R.~von Metzen, T.~Stieglitz, \href{https://www.nature.com/articles/s41378-022-00466-z}{A flexible protruding microelectrode array for neural interfacing in bioelectronic medicine}, Microsystems \& Nanoengineering 8~(1) (2022) 1--15, publisher: Nature Publishing Group.
\newblock \href {https://doi.org/10.1038/s41378-022-00466-z} {\path{doi:10.1038/s41378-022-00466-z}}.
\newline\urlprefix\url{https://www.nature.com/articles/s41378-022-00466-z}

\bibitem{ko_3d_2023}
D.~H. Ko, D.~Bates, H.~Karaosmanoglu, K.~Taredun, C.~Elton, L.~Jones, A.~Hosseini, A.~Partridge, {3D} microelectrode arrays, pushing the bounds of sensitivity toward a generic platform for point-of-care diagnostics, Biosensors \& Bioelectronics 227 (2023) 115154.
\newblock \href {https://doi.org/10.1016/j.bios.2023.115154} {\path{doi:10.1016/j.bios.2023.115154}}.

\bibitem{mateus_improved_2019}
J.~C. Mateus, C.~D.~F. Lopes, M.~Cerquido, L.~Leitão, D.~Leitão, S.~Cardoso, J.~Ventura, P.~Aguiar, Improved in vitro electrophysiology using {3D}-structured microelectrode arrays with a micro-mushrooms islets architecture capable of promoting topotaxis, Journal of Neural Engineering 16~(3) (2019) 036012.
\newblock \href {https://doi.org/10.1088/1741-2552/ab0b86} {\path{doi:10.1088/1741-2552/ab0b86}}.

\bibitem{muzzi_human-derived_2023}
L.~Muzzi, D.~Di~Lisa, M.~Falappa, S.~Pepe, A.~Maccione, L.~Pastorino, S.~Martinoia, M.~Frega, Human-{Derived} {Cortical} {Neurospheroids} {Coupled} to {Passive}, {High}-{Density} and {3D} {MEAs}: {A} {Valid} {Platform} for {Functional} {Tests}, Bioengineering (Basel, Switzerland) 10~(4) (2023) 449.
\newblock \href {https://doi.org/10.3390/bioengineering10040449} {\path{doi:10.3390/bioengineering10040449}}.

\bibitem{soscia_flexible_2020}
D.~A. Soscia, D.~Lam, A.~C. Tooker, H.~A. Enright, M.~Triplett, P.~Karande, S.~K.~G. Peters, A.~P. Sales, E.~K. Wheeler, N.~O. Fischer, A flexible 3-dimensional microelectrode array for in vitro brain models, Lab on a Chip 20~(5) (2020) 901--911.
\newblock \href {https://doi.org/10.1039/c9lc01148j} {\path{doi:10.1039/c9lc01148j}}.

\bibitem{Mller2015HighresolutionCM}
J.~M{\"u}ller, M.~Ballini, P.~Livi, Y.~Chen, M.~Radivojevic, A.~Shadmani, V.~Viswam, I.~L. Jones, M.~Fiscella, R.~Diggelmann, A.~Stettler, U.~Frey, D.~J. Bakkum, A.~Hierlemann, \href{https://api.semanticscholar.org/CorpusID:21743724}{High-resolution {CMOS MEA} platform to study neurons at subcellular, cellular, and network levels}, Lab on a chip 15 13 (2015) 2767--80.
\newline\urlprefix\url{https://api.semanticscholar.org/CorpusID:21743724}

\end{thebibliography}

\clearpage
\section*{Supplementary Information}
\label{suppinfo}
\renewcommand{\thefigure}{S\arabic{figure}} 
\setcounter{figure}{0} 
\renewcommand{\thetable}{S\arabic{table}} 
\setcounter{table}{0} 

\subsection*{S2.1.1 Equipment Required for Workspace}
\label{suppinfo:equipment}

\begin{itemize}
    \item \textit{\underline{Basic equipment}:} First, to handle neuron cultures, a fully functional biosafety cabinet of minimum biosafety level II (recommended by the Centers for Disease Control or CDC) with UV lights and air filters (HEPA) need to be adequately installed, along with an incubator with a supply of $5\%$ CO\textsubscript{2}, a centrifuge (preferably refrigerated) that supports tubes of various sizes/plates, a laminar flow hood, a water bath with proper temperature control, and finally, a fridge having both a $4^{\circ}$ Celsius refrigerator and a $-20 ^{\circ}$ Celsius freezer \cite{o2006setting, richmond2009biosafety}, as well as $-80 ^{\circ}$ freezers and liquid nitrogen-based storage for long-term storage of cells. One or more microscopes (fluorescent or nonfluorescent) is also essential for cell viability assessments. All institutes with biomedical and biological science departments have core facilities (mostly with a fee structure) that have these devices. For beginners, access to a cell culture core facility is the ideal stepping stone into the field. However, extra cautiousness is required to avoid cross-contamination in these shared facilities used by multiple groups.

    \item \textit{\underline{Consummables}:} After securing the basic equipment, the next step is to have enough cell culture disposables (i.e., treated/culture ready multiwell plates/dishes/flasks) to perform routine cell culture and sterile pipettes, micropipettes, centrifuge tubes, microtubes, and bottles to perform various mixing and transferring, and to store various reagents.

    \begin{itemize}
        \item[$\diamond$] \textit{\underline{Serological pipette and micropipette tips}:} Serological pipettes of different volumes are available in the market, ranging from $1$ml to $100$ml, suitable for dispensing needed volumes of liquid reagents with minimal effort. It is also important to have sterile micropipette tips for pipetting volumes ranging from $1$ul to $1000$ ul ($1$ ml). Suitable micropipettes are used for the precise dispensing of liquids and growth factors for neuron culture. Micropipettes are calibrated regularly by certified personnel.  
        
        \item[$\diamond$] \textit{\underline{Centrifuge tubes}:} For centrifuge tubes, it is important to have enough $1$ ml, $15$ ml, and $50$ ml tubes. There are various types of centrifuge tubes available in the market \cite{goodman2007centrifuge}. Smaller ones are useful for storing aliquots of growth factors and reagents to prevent contamination and degradation due to multiple freeze-thaw cycles. In comparison, larger ones are for processing/washing cells during centrifugation cycles and as temporary storage (short-term) of various reagents.
        
        \item[$\diamond$] \textit{\underline{Autoclavable Pyrex bottles}:} We would recommend to have at least $500$ ml and $1000$ ml Pyrex bottles. They can store large quantities of mixtures as well as distilled and autoclaved water. We would put special emphasis on having enough autoclaved water at hand, as it is crucial in every aspect of the culture. In many cases, having access to a water purification system for deionized water would be even more beneficial if we are working with liquids extremely sensitive to pH changes. It is important to sterilize water/homemade reagents by either autoclaving or filtering through $0.2\mu m$ filters.
        
        \item[$\diamond$] \textit{\underline{Multiwell plates}:} There are various types of cell culture wells available in the market, namely, $6$-well, $12$-well, $24$-well, $48$-well, and $96$-well. Each size has advantages and disadvantages. In our case, we mainly used $24$-well plates because the MEA devices we used have a culture-well size close to the individual wells in the $24$-well plates (see Section \ref{electro}). It is common to perform all the calculations for media volumes and concentrations based on the volumes of the wells.
        
        \item[$\diamond$] \textit{\underline{Petri dish}:} Other than the cell culture wells, we also use Petri dishes of various sizes. For our neuron cultures, we mainly used a $140$ mm Petri dish to act as containers for our MEA wells.

        \item[$\diamond$] \textit{\underline{Other consumables}:} It is important to have a steady supply of $70\%$ Ethanol, bleach, paper towels, Kim wipes, sterile nitrile gloves, lab coats (all required for keeping a sterile environment), pipette device filters, and CO\textsubscript{2} usage in the incubators. 

    \end{itemize}
\end{itemize}

\subsection*{S2.1.2 Environment Parameters}
\label{suppinfo:controlled_env}

\begin{itemize}
    \item \textit{\underline{Temperature}:} Mammalian cells are generally cultured at a temperature of $37^{\circ}$ Celsius, the average internal temperature of the human body. Most cell culture incubators are equipped with automated temperature control. However, it is always advised to use a conventional thermometer (digital or analog) to verify the chamber temperature regularly (recommended once every three months). Stable temperature maintenance is essential for normal metabolic activities and protein stability \cite{petrov_effect_2018, geneva_normal_2019}.
    
    \item \textit{\underline{CO\textsubscript{2} supply}:} The incubators are generally designed to have a constant supply of $5\%$ CO\textsubscript{2}. This concentration is selected to regulate the pH of the culture medium. The culture medium typically contains bicarbonate, which acts as a buffer. CO\textsubscript{2} interacts with water in the medium to form carbonic acid, which helps maintain the pH at a physiological level (around $7.2$ to $7.4$). Without proper CO\textsubscript{2} levels, the medium could become too alkaline or acidic, which could harm the cells \cite{ukhealth2020co2conc}. The CO\textsubscript{2} levels in the incubator must be confirmed and calibrated using either CO\textsubscript{2} Fyrite or digital CO\textsubscript{2} measurements.
    
    \item  \textit{\underline{Relative humidity}:} Neuron cells and the culture media are sensitive to humidity. It is crucial that the humidity is kept at a pretty high level ($\geq 95\%$ relative humidity) to ensure that the media does not evaporate. Maintaining high humidity ensures the stability of the osmotic balance, nutrient concentrations, and pH level \cite{mestres_factors_2021, ibidi2022humidity}. For standard CO\textsubscript{2} incubators, it is done by having a water basin filled with distilled and autoclaved water at the bottom of the incubator or, in the case of MEAs inside a Petri dish, using a smaller dish as a water chamber. It is crucial that we refill the water basin regularly with autoclaved water and ensure that we use water conditioner solutions such as Aqua-Clear (SP Bel-Art \ \#F$17093$-$0000$) to prevent the growth of algae, bacteria, molds, and biofilms. Also, it is vital that we do not use deionized water for metal basins, as it can react with the metal and damage the basin and other metallic components of the incubator over time.
    
\end{itemize}

\subsection*{S2.1.3.1 Primary Neuron Culture Factors}
\label{suppinfo:primary_neurons}

\begin{itemize}
        \item \textit{\underline{Age and location of harvested cells}:} The cells can be harvested at various developmental stages, from the embryonic state to postnatal, even adulthood. The age of the animal is selected based on the experiment requirements. For both C57 mice and Sprague-Dawley rats, E$18$ neurons are widely used, which are available commercially \cite{chakrabarti_defects_2007, dumanis_apoe_2011}. It is also relevant to decide on the brain region from which we want to harvest the cells. Cortical neurons, harvested from the cerebral cortex, are mostly associated with sensory perception, problem-solving, and decision-making. Hippocampal neurons from the hippocampus (a seahorse-shaped region of the brain) are more associated with memory storage and retrieval processes, learning, and emotions \cite{kandel_principles_2012}. Figure \ref{fig:e18_brain_atlas} shows the cerebral cortex and hippocampal regions in an E18.5 mouse brain atlas. For more information on the different brain regions at different points of mouse embryonic life, see References \cite{gilbert_chapter_2018, thompson_high-resolution_2014}. For this study, we used E$18$ embryonic mouse cortical neurons.

        \begin{figure}
            \centering
            \includegraphics[width=1\linewidth]{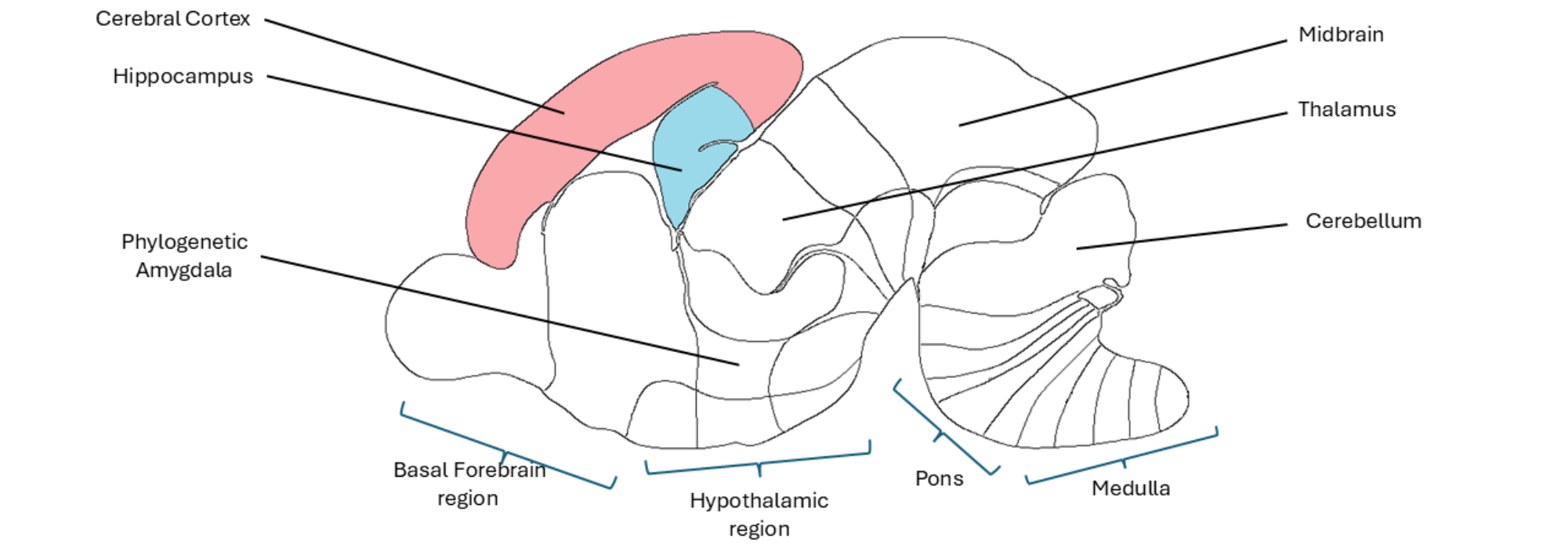}
            \caption{Cerebral cortex (pink) and hippocampal (cyan) regions, two major sources of the neurons used in primary culture, in an E18.5 mouse brain atlas (sagittal view).}
            \label{fig:e18_brain_atlas}
        \end{figure}

\item \textit{\underline{Waiting time before electrical activities}:} Although primary neuron cells have a limited lifespan in culture, they maintain essential characteristics of the cells \textit{in vivo}. It is easier for beginners to start with primary neuron culture because both pluripotent stem cells and neural stem cell-based culture protocols involve numerous additional culturing and maintenance steps and can be very complicated for new labs. Generally, it is possible to obtain good-quality electric signals from E$18$ cortical neurons within $11$-$18$ days \textit{in vitro} (DIV $11$-$18$) \cite{kamioka_spontaneous_1996, wagenaar_extremely_2006}. For our example data/study herein, we obtained good-quality electrical signals on DIV $14$. 

\item \textit{\underline{Cell procurement}:} There are multiple ways to obtain the cells required for primary culture. The easiest (but higher-cost) option is to buy the cells from commercial vendors. For this study, we purchased E18 mouse dissociated cortex cells (along with compatible media) from Transnetyx Tissue (Cordova, Tennesse $38016$, USA), formerly known as Brainbits, an established cell supplier in the USA (\url{https://tissue.transnetyx.com/}). The second option is to collaborate with labs that regularly maintain mouse colonies and can ensure a steady supply of neuron cells. Before proceeding, please check that the said laboratories have licenses and authorization from the Institutional Animal Care and Use Committee (IUCAC) \cite{curzer_three_2016}.

    \end{itemize}

\subsection*{S2.1.3.2 Cell Line-based Neuron Culture Factors}
\label{suppinfo:cell_line_factors}

    \begin{itemize}
        \item \textit{\underline{Cell line from cancer stem cells}:} 
        Immortalized glioblastoma and neuroblastoma samples are easy to maintain \textit{in vitro}, requiring minimal expertise. Importantly, these progenitors retain several properties similar to \textit{bona fide} neuronal progenitors \cite{bhaduri_outer_2020} and can be induced to electrophysiologically mature neurons with the addition of small molecules, such as retinoic acid \cite{kumar_data_2018}. However, these neurons acquire heterogeneous cell fates \cite{bhaduri_outer_2020}, meaning that although the neurons can be electrically mature, they may not exhibit the same behavior of a normally developed post-mitotic neuron and should be used cautiously. 
    
        \item \textit{\underline{Cell line from stem cells}:} Embryonic stem cells (ESCs) and induced pluripotent stem cells (IPSCs) derived cultures offer several advantages including but not limited to: (1) reproducibility of cell types \cite{velasco_individual_2019}, (2) a large toolkit of genetically modified and disease cell lines \url{https://www.wicell.org/} \cite{nowakowski_cerebral_2022}, (3) access to a wide range of cells derived from different species \cite{pollen_establishing_2019}, and (4) ability to generate and differentiate into cell types from all regions of the brain \cite{amin_generating_2023}. Human neuronal differentiation can give insights through disease modeling (i.e., culture cells derived from sources/tissues within specific genetic signatures associated with a disease) \cite{nowakowski_cerebral_2022} and may provide increased plasticity and function within SBI applications \textit{in vitro} \cite{goldwag_dishbrain_2023}. However, iPSC sources require long developmental timelines for maturation, require BSL2 grade equipment, and have a high level of oversight by the Institutional Stem Cell Research Oversight Committee (ISCRO)\cite{zettler_establishing_2007}. Other species, such as mice, can recapitulate similar cortical networks while maturing much more rapidly \cite{yoshiki_sasai_self-organization_nodate, elliott_internet-connected_2023}, producing extensive networks within 25 days, while human-derived cells can take 2-3 months. Here, we will attempt to lower the seemingly high skill barrier to entry by discussing how to culture 3D cortical brain organoids from mouse and human pluripotent stem cells. It is important to note that this is by no means a comprehensive guide on all possibilities of stem cell differentiation but a great starting point to begin work with stem cell-derived cultures.
        
    \begin{itemize}
        \item[$\diamond$] \textit{\underline{Common techniques for mouse and human cells}:}
        \begin{itemize}
            \item Both species require maintenance on 1--2\% v/v Matrigel Growth Factor Reduced (GFR) Basement Membrane Matrix (Corning \ \#$354230$), 1--2\% Geltrex\texttrademark{} LDEV-Free Reduced Growth Factor Basement Membrane Matrix (Thermo Fisher Scientific \ \#A$1413201$), or recombinant human protein vitronectin (Thermo Fisher Scientific \ \#A$14700$) coated plates. 
            \item If passaged, supplement media with $10$\,$\mu$M Rho Kinase Inhibitor (Y-27632, Tocris \ \# 1254) in the media for 1--2 days.
            \item Primocin (Invitrogen \ \#ant-pm-$05$) at $0.05$\,mg/mL is an antifungal, antibiotic, and anti-mycoplasma, making it preferred to prevent contamination.
        \end{itemize}
        \item[$\diamond$] \textit{\underline{Techniques only used for mouse cells}:}
        \begin{itemize}
            \item We recommend purchasing a feeder-free commercial mouse ESC line such as ES-E14TG2a (ATCC CRL-1821) to not have to deal with producing feeder cultures. Maintain in a GMEM base media with embryonic stem cell-qualified fetal bovine serum (Thermo Fisher Scientific \ \#$10439001$), $0.1$\,$\mu$M MEM nonessential amino acids (Thermo Fisher Scientific \ \#$11140050$), 1\,mM sodium pyruvate (Millipore Sigma \ \#S8636), 2\,mM glutamax supplement (Thermo Fisher Scientific \ \#$35050061$), $0.1$\,mM 2-mercaptoethanol (Millipore Sigma \ \#M3148), and $0.05$\,mg/mL primocin. The media needs to be supplemented with freshly added leukemia inhibitory factor (LIF) (Millipore Sigma \ \#ESG$1107$) and changed every day. 
        \end{itemize}
        \item[$\diamond$] \textit{\underline{Techniques only used for human cells}:}
        \begin{itemize}
            \item Maintain in StemFlex Medium (Thermo Fisher Scientific \ \#A$3349401$) or mTeSR\texttrademark{} Plus (StemCell Technologies  \ \#$100-0276$). Both have a stabilized FGF2-supplemented media, allowing the media to be changed every 2--3 days.  
        \end{itemize}
    \end{itemize}
    \end{itemize}

Differentiation into specified brain regions requires manipulation of morphogen pathways, typically through pharmacological perturbation, which is largely conserved through mammalian evolution, allowing for similar media compositions. However, the timelines of development and maturation are quite different \cite{eiraku_self-organized_2008, sasai_cytosystems_2013}. Focusing on cortical differentiation in 3D, we typically divide into 3 stages of media as shown in Table \ref{tab:cell-culture-schedule}: 
    
\begin{table}[htbp]
\centering
\small
\setlength{\tabcolsep}{4pt}
\begin{tabular}{|p{1.5cm}|p{2.5cm}|p{2.8cm}|p{2.8cm}|p{2.8cm}|}
\hline
& \makecell[c]{\textbf{Maintenance}\\\textbf{Media}} 
& \makecell[c]{\textbf{Cortical}\\\textbf{Differentiation}\\\textbf{Media}} 
& \makecell[c]{\textbf{Neuronal}\\\textbf{differentiation}\\\textbf{medium}} 
& \makecell[c]{\textbf{Neuronal}\\\textbf{maturation}\\\textbf{medium}} \\
\hline
Mouse & Day $-1$, seed 3K cells/well 
& Days $0$-5, change everyday 
& Days 5-14, change every 2-3 days 
& Days 14+, change every 2-3 days \\
\hline
Human & Day $-1$, seed $10$K cells/well 
& Days $0$-18, change everyday 
& Days 18-35, change every 2-3 days 
& Days $70$+, change every 2-3 days \\
\hline
\end{tabular}
\caption{Cell Culture Schedule for Mouse and Human Cells}
\label{tab:cell-culture-schedule}
\end{table}

\subsection*{S2.1.4.1 Coating Reagents}
\label{suppinfo:coating_reagents}
\begin{figure}
    \centering
    \includegraphics[width=0.75\linewidth]{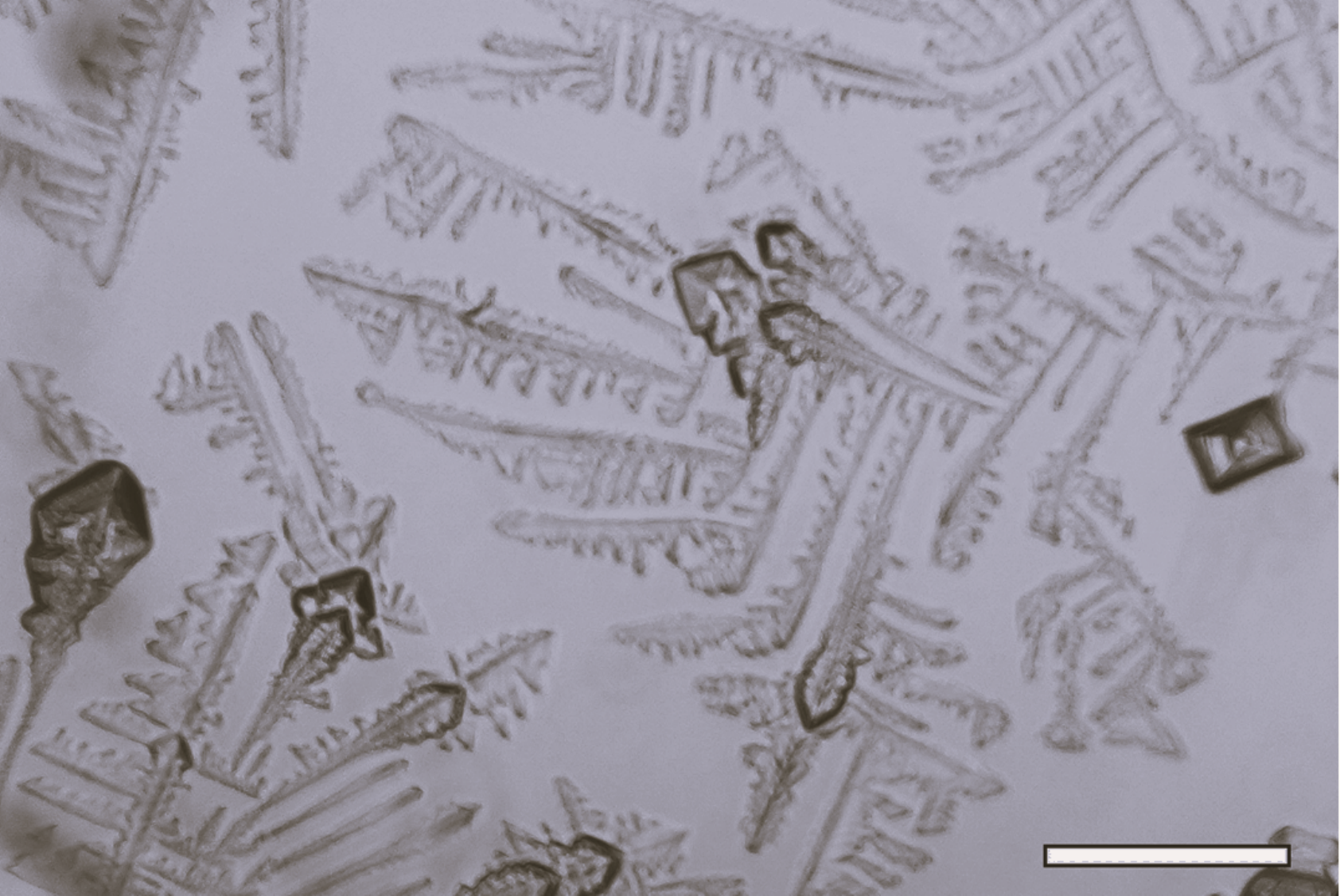}
    \caption{Excessive residual PDL forming crystalline structures that are visible even under standard microscopes. This excessive synthetic coating will result in a toxic environment in which no cell will be able to survive. Scale indicates $100\mu$m.}
    \label{fig:high_pdl}
\end{figure}

In our studies, we performed a primary coating with poly-d-lysine (PDL) (Corning\texttrademark \ BioCoat\texttrademark \ \#$354210$), followed by a secondary coating using Laminin (Corning\texttrademark \ \#$354239$). It is important to control the coating concentration in order to maintain one single molecular layer. Coating at high PDL or laminin concentrations will result in cell death, while low concentrations will result in limited cell attachment. This is especially true for high-density cell plating, where the cell density can be as high as $100,000$ to $400,000$ per $cm^2$, requiring adequate surface coating with adhesion molecules. Low cell adhesion will result in weaker signal acquisition in MEA wells and quicker detachment from the surface. It is also essential to wash the coating with deionized water or Phosphate-buffered saline or PBS (Corning \ \#$21040CV$) to ensure that only an appropriate amount of coating remains on the surface. Figure S2 shows an example of the effect of high PDL concentration during coating (Figure \ref{fig:high_pdl}), where the excess PDL forms crystalline growth, visible even under standard microscopes. These can also form with recommended PDL concentrations if the well is not aspirated and rinsed within the protocol’s time window. Ideally, the coating should be a single molecular layer and detectable only under electron microscopes. For this reason, it is important to follow the proper protocol to ensure proper coating and cell survival.

Following the MaxWell protocol, we prepared a final concentration of $0.1$mg/ml of PDL by diluting our procured $20$ mg PDL in $200$ ml of $1:20$ diluted $20\times$ Borate buffer (Thermo Scientific \ \#$28341$). We also prepared $0.02$ mg/ml of Laminin by mixing the Laminin with $50$ ml of NbActiv4 media (Transnetyx Tissue \ \#C$1477957$). After we prepare the solutions, we preserve them according to the protocol and reagent datasheets. We can prepare aliquots with the excess reagents and store them in $-20^{\circ}$C fridge. However, we need to ensure that we slowly re-thaw them overnight in the $4^{\circ}$C refrigerator to prevent any damage to the polymer structures when utilized, which might reduce the effectiveness of the coating.

\subsection*{S2.1.4.2 Primary Neuron Culture Medium}
\label{suppinfo:primary_medium}
Focusing on primary cell cultures, the initial stages of the culture require maintenance media that is rich in nutrients. Several options exist, including DMEM, GMEM, and Neurobasal media, all of which are generally supplemented with additional nutrients.  In our studies, we used NbActiv4 (Transnetyx Tissue \ \#NC$1477957$), which is a variation of Neurobasal media \cite{brewer_nbactiv4_2008}. To promote maturation, we switched to BrainPhys media, which is optimized for long-term electrophysiology \cite{bardy_neuronal_2015}. In our work, we  used BrainPhys Media\texttrademark \ (Stemcell Technologies \ \#$05790$) supplemented with B27 Plus (Gibco\texttrademark \ \#$A3582801$), N2 (Gibco\texttrademark \ \#$17502048$), and CD Lipid concentrate (Gibco\texttrademark \ \#$11905031$) \cite{elliott_internet-connected_2023}. 

Besides the recommended supplements, we added brain-derived neurotropic factor (BDNF) (Gibco\texttrademark \ \#PHC$7074$) in both NbActiv4 and Brainphys, which is crucial for cell survival and maturation \cite{ghosh_requirement_1994}. Following the datasheet, we reconstituted the $10\mu$g BDNF in $20\mu$L distilled water to obtain a concentration of $0.5$mg/mL. We further diluted it in $20$mL of $0.1\%$ bovine serum albumin (BSA) to obtain a final concentration of $100$ng/mL. To obtain the BSA solution, we dissolved $1$g BSA (Fisher Bioreagents \ \#BP671-1) into $1$L of PBS. Other than BDNF, we also added Primocin (InvivoGen \ \#ant-pm-1, \url{https://www.invivogen.com/primocin}), a less cytotoxic antimicrobial reagent, which provides the combined effect of antibacterial agents such as Penicillin/Streptomycin (Gibco\texttrademark \ \#$15140122$) or antifungal agents such as Amphotericin B, along with mycoplasma protection to some extent (Gibco\texttrademark \ \#$15290026$).

\subsection*{S2.1.4.3 Stem Cell Culture Medium for Neurons}
\label{suppinfo:stemcell_medium}
The media requirements for stem cell-based organoid culture process require different media preparation at different stages (initial plating and maintenance of stem cells, cortical differentiation, neuronal differentiation, and neuronal maturation) \cite{park_modulation_2024, quadrato_cell_2017}. We describe the media requirements below:

\begin{itemize}
    \item \textit{\underline{Maintenance medium}:} Passage cells and input respective number of cells per well in a lipidure-coated 96-well V-bottom plate. Maintain overnight in respective species maintenance media.
    
    \item \textit{\underline{Cortical differentiation medium}:} Glasgow minimum essential medium, 10\% knock-out serum replacement, 0.1 mM MEM nonessential amino acids, 1 mM sodium pyruvate, 2 mM Glutamax supplement, 0.1 mM 2-mercaptoethanol, and 0.05 mg/ml Primocin (Invitrogen \ \#ant-pm-05). Supplemented with Rho kinase inhibitor, WNT inhibitor (IWR1-$\varepsilon$, 3 $\mu$M, Cayman Chemical \ \#13659), and TGF-$\beta$ inhibitor (SB431542, Tocris \ \#1614, 5 $\mu$M).
    
    \item \textit{\underline{Neuronal differentiation medium}:} DMEM: nutrient mixture F-12 with GlutaMAX supplement (Thermo Fisher Scientific \ \#10565018), 1$\times$ N-2 supplement, 1$\times$ B-27 supplement minus vitamin A (Thermo Fisher Scientific \ \#12587010), 1$\times$ chemically defined lipid concentrate, and 0.05 mg/ml primocin.
    
    \item \textit{\underline{Neuronal maturation media}:} BrainPhys neuronal medium (Stem Cell Technologies \ \#05790), 1$\times$ N-2 supplement, 1$\times$ chemically defined lipid concentrate, 1$\times$ B-27 plus supplement (Thermo Fisher Scientific \ \#A3582801), 0.05 mg/ml primocin, and 0.5\% v/v Matrigel growth factor reduced (GFR) basement membrane matrix, LDEV-free.
\end{itemize}

\subsection*{S2.2.1 Methods for Cleaning Wells}
\label{suppinfo:cleaning}
\begin{itemize}
    \item \textit{\underline{Cleaning chemicals}:} Chemicals like $70$\% ethanol are easy to procure and can be easily used to clean non-sterile culture wells prior to use. However, in the case of MEAs, we need to check the datasheet and ensure that the MEAs will not get damaged from ethanol use. Also, we need to ensure that there is no residual ethanol left by properly aspirating it with distilled water and drying the wells. For this reason, a laminar flow hood can be extremely useful. If such a hood is not available, using a biosafety cabinet with air circulation on can also be a feasible option.

    1\% Tergazyme (Electron Microscopy Sciences \ \#$60520$), EDTA (Invitrogen \ \#$AM9260G$), or Collagenase (Sigma-Aldrich \ \#$C0130$) are reagents that break down enzymes and collagens, and other large molecules to keep MEAs free off any existing organic contaminants. However, we need to thoroughly wash the wells using distilled, deionized water or PBS after using these reagents; otherwise, any residual amount will result in apoptosis. After washing the wells, it can be dried using the above-mentioned approaches.

    In our case, we followed MaxWell Biosystems' protocol, which requires both $1$\% Tergazyme and $70$\% Ethanol.

    \item \textit{\underline{Disinfecting equipment}:} Low-pressure plasma cleaning devices are effective in cleaning the wells \cite{petasch_low-pressure_1997}. Also, short-term UV radiation is a feasible approach in conjunction with other methods to kill any residual microorganisms in standard wells. However, we must refrain from using UV light to sterilize HD-MEA wells as it might severely damage them. Since most biosafety cabinets already have a UV light, beginner groups can leave their culture wells inside the incubator. However, it is important to follow the manufacturer's recommendation for exposure time. Otherwise, the material structure of the culture wells can get damaged due to an exceeding amount of radiation. Generally, wells are not autoclave-able, so we would not recommend this approach. As MEAs, especially CMOS-based HD-MEAs like MaxOne, are far more sensitive, we recommend cleaning using only chemicals following the manufacturer's suggested protocol.
\end{itemize}

Occasionally, the cleaning process might reduce or destroy the hydrophilicity of the MEA or well surface, turning it into a hydrophobic one. In that case, please follow the proper protocols provided by the manufacturer to restore the surface to hydrophilicity.

\subsection*{S2.2.2 Key Factors for Coating Wells}
\label{suppinfo:coating_factors}
\begin{itemize}
    \item \textit{\underline{Coating preparation}:} Before we apply the coating, we must prepare both the primary and secondary coating reagents. We prepare the primary coating (PDL in our case) solution either by dissolving it in deionized water or a buffer solution to maintain a stable pH value. For the secondary coating, we dissolve the biopolymers, such as Laminin, in the culture medium we intend to use. This ensures that the coating and the media together will closely emulate the ECM in a living tissue.

    \item \textit{\underline{Schedule management}:} We need to properly plan the coating schedules for both the primary and the secondary coating. We consider not only the time it would take for the aliquots to thaw overnight slowly but also the time required to ensure proper coating, rinsing, aspiration, and drying, as well as the time for secondary coating. It must be mentioned that while incubating the well after the secondary coating, we need to simultaneously prepare the cells for plating to minimize any shock related to temperature change. We have discussed this further in Subsection \ref{sec:culture_protocol}.

    \item \textit{\underline{Evaporation prevention}:} When we incubate for both the primary and the secondary coating, we must ensure that the coatings do not dry up in any way. Ideally, we intend to coat only the electrode region at the center (for adequate cell density in the recording region), which requires a small volume of liquid solution that can easily evaporate without special care. If we incubate it inside the incubator, we need to ensure the water basin has enough water to prevent evaporation. If we incubate it outside at room temperature, it is better to keep it in a sealed chamber with a water chamber that prevents evaporation.   
\end{itemize}   

\subsection*{S2.2.3 Key Factors for Plating Cells}
\label{suppinfo:plating}
\begin{itemize}
    \item \textit{\underline{Prior preparation}:} Preparation for cell plating is something we need to start in parallel while we are incubating for the secondary coating, or even beforehand, depending on the state of the procured cells (elaborated in Subsection \ref{sec:neuron_culture_aspects}). Prior preparation generally involves temperature equilibration, dissociation (if non-dissociated cells are procured), and separation using a centrifuge device. At every step of introducing various reagents to the cells, we must ensure that the cells and the added reagents are at the same temperature. A large temperature difference will shock all the cells, resulting in apoptosis before we plate them. 
    
    \item \textit{\underline{Use of centrifuge device}:} After the cells are dissociated, we separate the cells from the storage media using a centrifuge machine at the rotation speed and spin times recommended by the cell vendors. Generally, neurons are larger and more delicate than other cells. Thus, a lower speed (around $200$G force) and a shorter time (around $1$ minute) are sufficient to separate the cells without any damage \cite{brewer_optimized_1993}. High speeds and long spin times will damage cells.

    \item \textit{\underline{Cell counting before plating}:} After we separate the cells using a centrifuge device, we need to count the cells before plating them. We can take a small sample from the suspended cells and count the cell density using a hematocytometer. If needed, we can also take multiple small samples and compute the average density to obtain a more accurate representation of the number of cells. If we are further interested in cell viability, we can use Trypan Blue (Gibco\texttrademark \ \#$15250061$), a staining agent, to count the number of alive cells vs dead cells in a sample. It must be mentioned that Trypan Blue is toxic to the cells and, therefore, only used for terminal experiments. This is the reason we can use it on a separate small sample and not on the main cell suspension.

    \item \textit{\underline{Plating the cells}:} Once the desired cell density is achieved, we use a micropipette to precisely plate cells onto the coated region. While plating the cells, we occasionally swirl the tube gently to prevent sedimentation. After plating, we should avoid adding more media immediately, which could disrupt cell attachment. Instead, we incubate a small volume for $20$-$30$ minutes to allow cell attachment before the media evaporates. Once attached, we can add an excess of warmed, CO\textsubscript{2}-equilibrated media and return to the incubator. However, we should be mindful not to incubate for more extended periods, as it can dry out the media and harm the cells.
    
\end{itemize}

\subsection*{S2.2.4 Key Factors for Media Change}
\label{suppinfo:feeding}
\begin{itemize}
    \item \textit{\underline{First change on DIV 1}:} It is recommended to perform the first media change within $24$ hours of plating. Generally, after plating the cells, it is expected that there will be some cell debris from the dissection process, as well as a fair amount of dead cells caused by chemical and mechanical factors (both before and after plating) that result in the failure of some cells to attach effectively. This first media change allows the removal of dead cells from the environment, not only preventing them from creating a toxic environment but also allowing the living neurons enough space to grow and form synaptic connections.

    \item \textit{\underline{Gently changing the media}:} After this first media change, we change the media frequently. As mentioned in Subsection \ref{sec:coating_and_medium}, we only perform a partial media change (generally $50\%$) instead of changing the whole volume of culture media. This minimizes potential mechanical and osmotic stress and also helps maintain biochemical factors that are produced by the cells themselves in the culture media. To perform a media exchange, we gently pipette half of each well's liquid contents from the boundaries of the wall so as not to touch or stress the cells. Tilting the wells a bit (not to the point of exposing the cells) could help guide the debris to the side, which we can remove more easily by pipetting.

    \item \textit{\underline{Adapting the media change routine}:} How frequently we change the media depends on the number of cells. The higher the number of cells, the higher the metabolic activity and the more frequently we need to do media change. For adequately high cell concentrations with high levels of electrical activity, we can change the media three times a week. However, these frequent media changes put a strain on the cells, which might reduce cell longevity. We need to monitor any changes in pH values (and thus changes in media color) and take appropriate responses. If the media starts turning yellowish (acidic), we replace the media. If the media turns pinkish (basic), we check the water basin and other factors causing evaporation.

    \item \textit{\underline{Switching from maintenance media to maturation media}:} As mentioned in Subsection \ref{sec:coating_and_medium}, we need to change from a maintenance media (such as NbActiv4) to a maturation media (such as BrainPhys) after a week. Instead of changing the whole media from NbActiv4 to BrainPhys, we recommend doing it over multiple days by performing partial changes. In this way, the cells would not suddenly experience a change in the nutrient levels and could adapt slowly. We can start changing the media from DIV $7$. By DIV $10$, we will mostly have BrainPhys media in the wells. Then, the cells will become active around DIV $11-14$.
    
    \item \textit{\underline{Regularly checking cell health}:} Although we recommend mixing antibacterial and antifungal agents (such as Primocin), we must watch for any contamination. It is recommended that cell health be checked under a microscope to see any issues with cell growth and network formation. If the MEA wells are transparent, an ordinary transmitting microscope is enough. Otherwise, we need to use a reflective light microscope. We discussed this in more detail in Subsection \ref{sec:cell_assess}.
    
\end{itemize}

\subsection*{S2.4 Reasons behind Contamination}
\label{suppinfo:contamination_reasons}
\begin{enumerate}
    \item \textit{\underline{Biosafety cabinet being unsterile}:} This situation can be prevented by turning on the UV light inside the biosafety cabinet for 10-15 minutes before use. After turning it off, switch on the airflow and raise the glass to its designated height (an alarm will sound if it exceeds the limit). We then need to wipe the interior with 70\% ethanol or bleach wipes (if needed). Additionally, weekly deep cleaning can be done using antifungal sprays.

    \item \textit{\underline{Cells exposed to airborne contaminants while moving them around}:} To prevent this from happening, we should always transport cultures in a sealed, disinfected carrier box or seal culture plates and tubes with parafilm. MEA wells should be placed in a sealed Petri dish to prevent evaporation. We must take special care when placing sterile items in the refrigerator, as inactive contaminants may be present. Similar precautions are needed for the incubator as well, especially when we put anything inside of it.

    \item \textit{\underline{Microbes getting inside the biosafety cabinet through equipment from outside}:} Even with precautions, contaminants can enter through the surfaces of gloves, lab coats, and equipment. To minimize this, we need to spray all non-paper items with 70\% ethanol, paying extra attention to the bottoms. Trash containers should be cleaned with bleach beforehand. We must open the packages containing the pipette tips and culture plates near the cabinet's opening and bring the contents inside without the external cover. We must also use fresh gloves, replace lab coats as recommended, and avoid blocking air vents. When pipetting protein solutions, please avoid bubbles, as they create anaerobic microenvironments conducive to microbial growth.

    Mycoplasma (type of bacteria) contamination is another serious concern in mammalian cell culture \cite{drexler_mycoplasma_2002}. Mycoplasma contamination usually does not get noticed because of a lack of visible symptoms in the cells. However, they affect several aspects of the physiology of the cells, thereby compromising the results. To minimize this, proper aseptic culture conditions should be met strictly. Also, it is highly recommended to test cell culture samples once every three months for mycoplasma contamination. Mycoplasma contamination can be detected using commercial kits, such as the Universal Mycoplasma Detection Kit, a PCR-based kit (ATCC \ \#$30-1012$K) for mycoplasma detection in cell culture.

    \item \textit{\underline{Reagents and consumables past their expiration date}:} Finally, please do not use any expired reagent or consumable equipment for any cell culture. Aliquotes can be prepared for long-term storage, and proper storage instructions should be followed, as mentioned in Subsection \ref{sec:logistics}, to maximize their utility and shelf life. However, if we are doubtful of any reagents (suspicious color, cloudiness, smell, any unwanted particle or speckle), it is always better not to take any risks and throw them away. We must always check the condition of the reagent before using it.
\end{enumerate}

\subsection*{S2.5 Key Factors for Proper Logistics}
\label{suppinfo:logistics}
\begin{itemize}
    \item \textit{\underline{Procuring and storing based on shelf life}:} Since many cell culture reagents have short shelf lives, we need to ensure a steady supply. Simpler reagents like cleaning agents ($70\%$ ethanol or bleach), distilled water, and buffer solutions (PBS, Borate) have longer shelf lives and can be stored at room temperature if kept sterile, and bulk procurement is fine. However, most culture-specific reagents, like media (NbActiv4, BrainPhys), supplements (such as B27, N2, CD Lipid concentrate, etc.), biological or synthetic polymers (such as PDL, PEI, POI, Laminin, Matrigel, etc.), proteins and enzymes (such as Papain, Tripsin, Albumin, Fetal bovine serum, etc.), antibiotic and antifungal agents (Penicillin/Streptomycin or Primocin), nonfluorescent and fluorescent stains (such as Trypan Blue, Fluorescein diacetate, Calcein AM, Propidium iodide, etc.) require cold storage (4°C, -20°C, or -80°C). We should follow storage guidelines to extend shelf life and prevent contamination.

    \item \textit{\underline{Preparing aliquots for greater efficiency}:} For reagents that require repeated use or reagents stored in subzero storage temperature, it is recommended not to store the whole bottle but to prepare aliquots of the reagents (i.e., divide them into smaller portions) and then store them separately. It is never a good idea to thaw and freeze any reagent repeatedly, as it can cause damage over time and might harm the cell culture. The number and amount of aliquot depends on how much reagent is required for one session of cell culture and the reagent's shelf life. It is an excellent practice to do all the calculations beforehand before following any cell culture protocol.

    \item \textit{\underline{Maintaining inventory of necessary apparatus and consumables}:} We recommend keeping proper tabs on the number of pipettes and centrifuge tubes available, as they are essential for transferring and storing reagents. Although we could borrow them from other laboratories or share the core facility resources, it is better to ensure that we are aware of our inventory and are prepared before and during the cell culture process.
    
\end{itemize}

\subsection*{S3.1 Neuron Signal Recording Techniques}
\label{suppinfo:neuron_recording}
\begin{enumerate}
    \item \textit{\underline{Intracellular recording}:} Intracellular recording refers to the measurement of electrical signals from inside a neuron or other cell. This technique allows researchers to measure the membrane potential, including detailed dynamics of action potentials and subthreshold activities, such as synaptic potentials. There are various technologies available for this purpose:
    \begin{enumerate}
        \item \textit{\underline{Patch clamp}:} Patch clamp techniques allow for precise measurements of the membrane potential and ion channel activities in single neurons \cite{nam_vitro_2011}. In the whole-cell configuration, researchers can directly access the intracellular space, while cell-attached configurations measure ion channel activities without penetrating the cell body. Although patch clamping provides valuable insights into synaptic inputs, such as excitatory and inhibitory postsynaptic currents (EPSC and IPSC), which reflect network activities, its low throughput makes it less feasible for large-scale network analysis. 
        
        \item \textit{\underline{Calcium imaging}:} Calcium imaging allows us to observe neuron activities by measuring calcium ion concentrations in cells using fluorescence imaging in conjunction with calcium-sensitive dyes. Calcium plays a critical role in intracellular signaling and is highly correlated with neuronal activities. Hence, calcium imaging provides us with an indirect and non-invasive measurement of electrophysiological activities in a neural network. However, since this method relies on the influx of calcium ions into cells following neuronal activities, this process is slower than the actual electrical signals (action potentials), resulting in lower temporal resolution. It has a high throughput and allows population-level recording, unlike patch clamps. Calcium imaging can involve either the use of calcium-sensitive dyes like Fluo-4 or Fluo-8AM \cite{cameron_prolonged_2017}, which fluoresce upon binding to calcium ions, or genetically encoded calcium indicators (GECIs) such as GCaMP. In the case of GECIs, we can use adeno-associated virus (AAV) vectors to infect the neuron cells with specific viruses to modify their genes to perform certain tasks and, in our case, express the calcium indicator protein. For example, we can use an AAV, which carries a gene for a calcium indicator like gCaMP, which fluoresces in the presence of calcium \cite{grodem_updated_2023}. We also need a promoter, such as the  hSYN1 (human Synapsin 1, which is also suitable for imaging rodent neurons), which drives gene expression in synaptically active neurons, ensuring that the calcium indicator is expressed in the neurons we want to monitor. However, the use and disposal of AAVs might require stricter biosafety clearance and authorization from relevant authorities, so not all core research facilities might have the necessary permission. It is essential to discuss this with the core directors before proceeding with ordering AAVs from vendors such as Addgene (for example, we can use AAV-hSyn1-GCaMP6f-P2A-nls-dTomato after signing the necessary material transfer agreements (MTA); \url{https://www.addgene.org/51085/}). 
        
        \item \textit{\underline{Voltage imaging}:} Voltage imaging is another intracellular and non-invasive imaging method that allows us to monitor changes in neuronal membrane potential in real-time using voltage-sensitive dyes or genetically encoded voltage indicators (GEVIs) such as ArcLight or QuasAr2 \cite{antic_voltage_2016, adam_voltage_2019}. These indicators change their fluorescence in response to shifts in membrane potential, providing us with a direct measure of electrical activities. Unlike calcium imaging, voltage imaging has a higher temporal resolution of action potentials and subthreshold voltage dynamics, making it valuable for studying rapid neural processes. To implement this, we can use viral vectors like AAVs to express GEVIs in neurons. These indicators can be targeted to specific cells using promoters like hSYN1, ensuring that the voltage signals are detected from the appropriate neuronal populations. Although voltage imaging provides high temporal resolution, it suffers from a lower signal-to-noise ratio compared to calcium imaging, making it technically challenging. Additionally, GEVIs and voltage-sensitive dyes often require sensitive imaging systems, such as high-speed cameras with high frame rates and powerful light sources. As with calcium imaging, the use of viral vectors for GEVI expression requires biosafety clearance from relevant authorities and must be discussed before ordering them.
    \end{enumerate}

    \item \textit{\underline{Extracellular recording}:} Extracellular recording involves placing an electrode outside a cell, usually in close proximity to the neurons of interest. Instead of measuring the membrane potential, this technique detects changes in the electric field generated by the collective activities of many neurons or by action potentials (spikes) from individual neurons. There are mainly two ways for extracellular recording:
    \begin{enumerate}
       \item \textit{\underline{Neural probes}:} It is feasible to obtain recordings from a small population of neurons by using neural probes \cite{jun_fully_2017}, such as Neuropixel by IMEC (Kapeldreef 75, Leuven $3001$, Belgium; \url{https://www.neuropixels.org/}). The primary advantage of using these probes is their flexibility. It is possible to grow our neuron cells in a standard Petri dish or multi-well setup and then insert the neural probes directly to a desired recording location. This is adequate if we only wish to focus on sparse locations in the neuron culture. Currently, studies are being done to increase the biocompatibility of the electrodes to ensure long-term recording \cite{gao_nanotechnology_2021, hong_novel_2019}.

        \item \textit{\underline{MEA wells}:} Another option is an \textit{in vitro} microelectrode array well, where neuron cells are cultured on a surface integrated with a multi-electrode array \cite{nam_vitro_2011}. Although it is not as high resolution as a neural probe, MEA wells allow us to record simultaneously from a larger network region. Compared to neural probes, MEA wells are less sensitive to single-cell events. They are better for detecting overall interactions between neurons in the culture, which is the primary interest in the development of SBI. Since MEA wells are less invasive, they are also more suitable for long-term recording. For this reason, our group focused on MEA recordings and data presentation herein. 
    \end{enumerate}
\end{enumerate}

\subsection*{S3.2 Key Characteristics of MEA Wells}
\label{suppinfo:MEA_char}
There are various types of MEAs available based on the culture requirements. Suppliers such as Multichannel Systems (MCS), Axion Biosystems, and Alpha MED Scientific (Ibaraki, Osaka $567$-$0085$, Japan; \url{https://www.med64.com/}) provide predominately planar electrode array MEA. Other suppliers such as MaxWell Biosystems and 3Brain AG ($8808$ Pfäffikon, Switzerland; \url{https://www.3brain.com/}) focus on CMOS MEA. Existing commercially accessible MEA typically do not use penetrating electrodes (which can acquire signals from above the culture surface), although exceptions such as 3Brain AG do provide partially penetrating electrodes. Significant works on a range of other MEA systems are underway and may increase the optionality in the future \cite{sileo_electrical_2013, steins_flexible_2022, ko_3d_2023, mateus_improved_2019, muzzi_human-derived_2023, soscia_flexible_2020}. 

\begin{itemize}
\item \textit{\underline{Electrode technology and material}:} MEAs commonly use high-conductivity, biocompatible materials such as platinum, gold, and titanium for their electrodes. These materials are chosen for their stability, durability, and low impedance, all of which are essential for neural signal recording. Traditional MEAs, such as those from MCS and Axion BioSystems, utilize these materials to create stable and long-lasting electrodes capable of capturing extracellular signals from neural networks. However, these MEAs are often limited by their relatively low electrode density and poor spatial resolution. Additionally, the absence of integrated on-chip electronics can make signal processing less efficient, as external amplification and multiplexing are required. In contrast, newer CMOS-based high-density MEAs, such as those developed by Maxwell Biosystems and 3Brain, as well as the CMOS-MEA$5000$-System from MCS, provide much higher electrode density and spatial resolution. The CMOS technology allows for the integration of dense electrode arrays with on-chip electronic components like amplifiers and multiplexers, which improve signal processing and reduce noise. In our work, we have utilized the MaxOne HD-MEAs from Maxwell Biosystems, which employ platinum-based high-density CMOS electrodes for precise high-resolution recordings \cite{Mller2015HighresolutionCM}.

\item \textit{\underline{Electrode size and density}:} Electrode size and density vary significantly across manufacturers to meet different research needs. Standard and high-density options are available from most companies. Standard MEAs do not provide as high resolution as those using CMOS technologies. For example, MCS  $60$MEA$200/30$iR-Ti MEAs have an inter-electrode distance of $200 \mu m$, a sensing area of $1.4 \times 1.4 mm^2$, containing $59$ electrodes (arranged in $8 \times 8$ grid, resulting in density of $32.65$ electrodes$/mm^2$). On the other hand, MaxOne HD-MEAs have an inter-electrode distance of $17.5 \mu m$, a sensing area of $3.85 \times 2.10 mm^2$, containing $26,400$ electrodes ($3,265$ electrodes$/mm^2$). Thus, the MaxOne HD-MEAs should have at least $100$ times the resolution as compared to the MCS MEAs.

\item \textit{\underline{Surface material and hydrophilicity}:} For standard MEAs made by MCS and Axion BioSystems, glass is the most commonly used substrate. For high-density arrays, companies like MaxWell Biosystems and 3Brain use CMOS substrates built on a silicon base. In most cases, MEA surfaces are hydrophobic by default, requiring treatments before cell plating to turn the surfaces hydrophilic, which is necessary for cell attachment. For example, MCS MEAs require pre-conditioning by plasma cleaning, protein pre-coating using fetal bovine serum (FBS) (Gibco\texttrademark \ \#$A3160401$), BSA, or simply the culture media. MaxOne requires pre-conditioning by incubating with the culture media for two days.

\item \textit{\underline{Cell viability assessment}:} As previously discussed in Subsection \ref{sec:cell_assess}, standard MEAs are generally transparent, while HD-MEAs are generally opaque. Thus, we must prepare our microscopy setup accordingly before choosing our MEAs.

\item \textit{\underline{Background noise level}:} Since the recorded extracellular electrical activities generally vary in the $50-500 \mu$V range, we need the MEAs to be as free from noise as possible. According to the datasheets of the MaxOne HD-MEAs, the background noise is within $\pm 5 \mu V$, which we found accurate for our MEA wells. Had we found any higher noise in any electrode, we would have discarded its recorded signals, which we did not need to do in our example experiments presented herein.

\item \textit{\underline{Re-usability}:} Most MEAs are reusable, provided they are maintained and cleaned according to manufacturer guidelines. MaxOne HD-MEAs can also be used multiple times after cleaning, but the electrodes will lose their effectiveness over time, and new MEAs must be procured once background noise exceeds acceptable limits following multiple uses. It is also possible to set up a regular supply schedule with the companies or buy in bulk. Still, it needs to be done after logistical consideration for cost-effectiveness, as discussed in Subsection \ref{sec:logistics}.

\end{itemize}

\subsection*{S3.3 Additional Electrical Activity Information}
\label{suppinfo:electrical_activity}

\begin{figure}[ht!]
    \centering
   \includegraphics[width=1\linewidth]{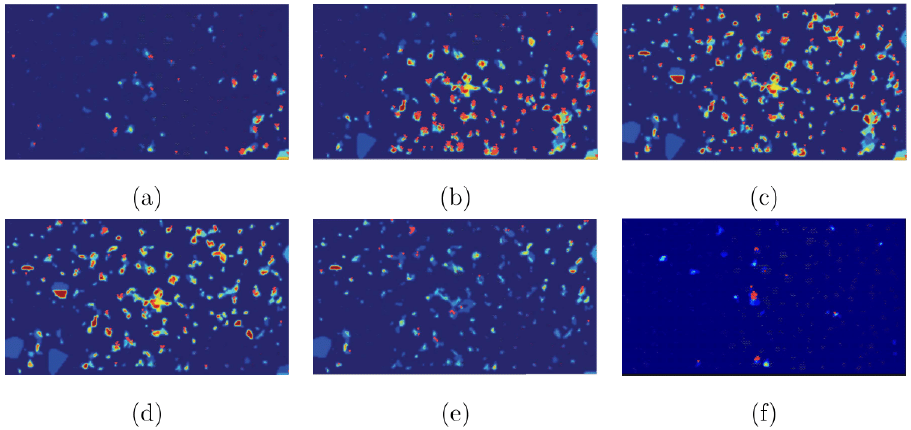}
    \caption{Multiple snapshots of a single burst activity in chip $3$. Here, we can see the synchronous burst activity originating from the bottom-right corner and propagating toward the top-left corner. The activity sustained for approximately 1 second.}
    \label{fig:burst_screenshot}
\end{figure}

We show six chronological snapshots of one burst activity in chip $3$ in Figure \ref{fig:burst_screenshot} shown as an example of electrical activity on a MaxOne HD MEA. Clearly, the burst originates from the bottom-right corner and propagates through the network towards the top-left in the chip, all within less than a second. This shows the interconnectivity and synchronous activity of the cultured neural network.

\begin{figure}[ht!]
    \centering
    \includegraphics[width=1\linewidth]{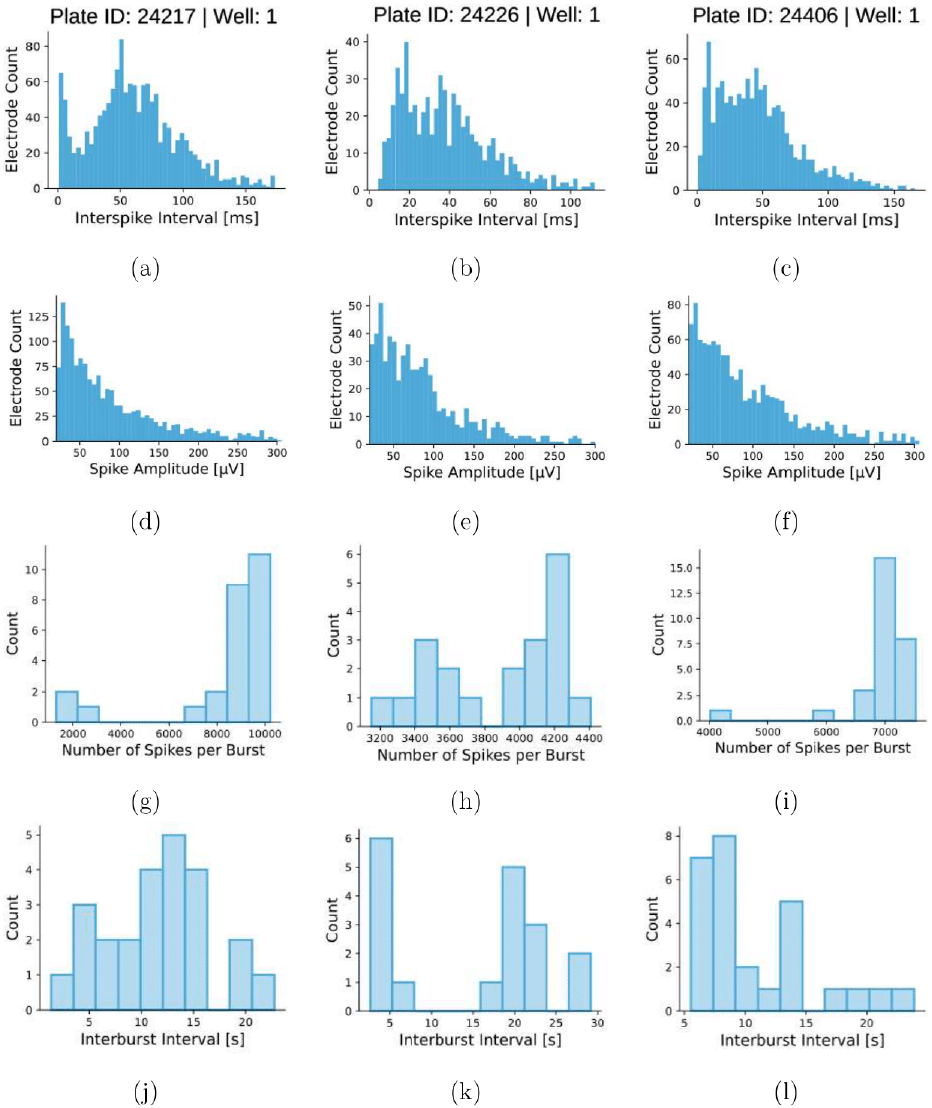}
    \caption{Histograms of interspike intervals (ms), spike amplitudes ($\mu V$), the number of spikes per burst (only for detected synchronous bursts), and synchronous burst intervals from all three chips. The columns represent the data from chips $1$-$3$ respectively.}
    \label{fig:activity_histograms}
\end{figure}

Moreover, Figures \ref{fig:activity_histograms}(a)-(f) show the histograms of the interspike intervals (which is the inverse of the firing rate) and the spike amplitudes of three chips we cultured during our trial experiments. The distributions of the interspike intervals roughly follow the same shape, the only difference being the high-frequency activities ($<10 ms$) in the chips $1$ and $3$, present in the large neural clusters.

\end{document}